\documentclass[12pt,twoside,a4paper,amsmath,amssymb,apsmy,showkeys,showpacs,nofootinbib]{revtex4}

\usepackage[cp1251]{inputenc}
\usepackage[russian,english]{babel}
\usepackage[left=2.5cm, right=2.5cm, top=2.5cm, bottom=2.2cm, bindingoffset=0cm]{geometry}
\usepackage[T2A]{fontenc}
\usepackage{subfig}
\usepackage{graphicx}
\usepackage{graphics}
\usepackage{epsfig}
\usepackage{float}
\usepackage{multirow}
\usepackage{bigstrut}
\usepackage{latexsym}
\usepackage{array}
\usepackage{dcolumn}
\usepackage{bm}

\begin{document}

\thispagestyle{myheadings}

\title{Multiparticle production in electron-positron annihilation 
}

\author{E.S. Kokoulina}
\email{kokoulina@jinr.ru}
\affiliation{Joint Institute for Nuclear Research, Joliot-Curie str., 6 
	Dubna, Moscow region, Russia, 141980 and
	and GSTU, Prospect Octiabria, 48,
	Gomel, Belarus, 246746 }

\begin{abstract}
Multiparticle production in hadron and lepton interactions still attracts our attention. Simulation by using Monte Carlo event generators is performed before planning any experiment. But it often overestimates (or underestimates) experimental data. These generators are based on the  theory of strong interactions, quantum chromodynamics (QCD), which is capable of performing calculations only in the perturbation theory. Soft processes that make up a significant contribution in high-energy interactions are forced to involve phenomenological models. 

Of all multiparticle production processes, electron-positron annihilation is the theoretically cleanest, proceeding via an intermediate virtual photon or $Z^0$-boson followed by quark-antiquark pair creation. QCD describes well the development of quark-gluon ($qg$) cascade as marcovian branching process, that is called first stage. 

The transformation of quarks and gluons produced in the $qg$-cascade into observable hadrons occurs in the second stage, hadronization, to which perturbation theory is no longer applicable. The choice of a scheme for it is based on experimental data. 

Convolution of $qg$-cascade and hadronization allowed us to describe the multiplicity in practice all processes of multiple production in both lepton and hadron high-energy collisions. This model is called the gluon dominance model.

Several decades have passed since a series of $e^+e^-$ annihilation experiments were carried out. Now, the main interests of high energy physicists are focused on the study of multiparticle production  in proton and heavy ion collisions. Their research revealed many new results in the theory of strong interactions, including the hadronization. That is why it appeared necessary to analyze multiplicity n $e^+e^-$-annihilation again.

\end{abstract}


\maketitle

\section{Introduction}
Multiparticle production (MP) of secondary particles in high energy interactions accompanies any experiment. Their study began in cosmic rays and continues at modern accelerators in interactions of protons, heavy ions and at electron-positron colliders. Over the past 50 years, the energy of colliding particles has been increased to the TeV region. Physicists are developing new technologies that will increase energies to hundreds of TeV.

The number of secondary particles formed in a collision is called multiplicity.
A distinction is made between the multiplicities of charged, neutral particles, and the total multiplicity (their sum). The number of secondary particles produced in a single interaction varies randomly from one event to another. Their average multiplicity increases with energy approximately logarithmically.

Research of MP stimulated the development of the theory of strong interactions, quantum chromodynamics (QCD) \cite{QCD}, 
which allows us to calculate the characteristics in the region of large momentum transfers, where the constant of strong interaction is small and perturbation theory (PT) \cite{QCD1} is applicable. QCD speaks the language of quarks and gluons, which are not directly observed in experiments. The transition from quarks and gluons to observable hadrons in this region of QCD is difficult.

To pass from quarks and gluons to experimentally observed hadrons, phenomenological models are developed \cite{Model1,Model2,Model3,GDM1,TSM,GDM2,GDM3,GDM4}, and rather complex time-consuming lattice calculations are performed. Before any experiment, physicists simulate the operation of their setup using Monte Carlo generators. They allow predictions to be made when planning an experiment.

At the same time, it often turns out that the results of modeling differ significantly from the experimental ones. This forces us to make changes, or, as physicists say, to adjust the generators to data. And so at each increase of energy. A particularly large discrepancy is observed in the region of high multiplicity, which significantly exceeds its average value.

The gluon dominance model (GDM) was originally developed to study the MP in electron-positron annihilation processes \cite{GDM1,TSM}. This model is a convolution of two stages. The first stage, called the quark-gluon cascade, is described by the differential-difference equations \cite{KUV,Giov} as a Markov branching process. The cascade develops due to the fission of gluons $g\to gg$ and the bremsstrahlung of the gluon $q\to qg$ by the quark, described by PT QCD \cite{KUV,Giov}.

To describe the second stage, hadronization, a phenomenological scheme is used. Its choice is based on the experimental behavior of the second correlation moment \cite{Rush}
\begin{equation}
	\label{eq}
	f_2 = D_2 -\overline n = \overline {n(n-1)} -\overline n^2,
\end{equation}
where $D_2$ and $\overline n$ are variance and the average multiplicity, the upper bar means averaging. $f_2$ changes sign from negative at low energies to positive at high. In addition, the region of negative values of $f_2$ differs significantly for proton collisions and annihilation processes ($e^+e^-$, $p\overline p$, $K^-p$). Experimental values of $f_2$ for average multiplicity of negative charged particles are presented in Figure 1 \cite{Rush}.
\begin{figure}[H]
	\leavevmode
	\centering
	\includegraphics[angle=0, width=0.45\textwidth]{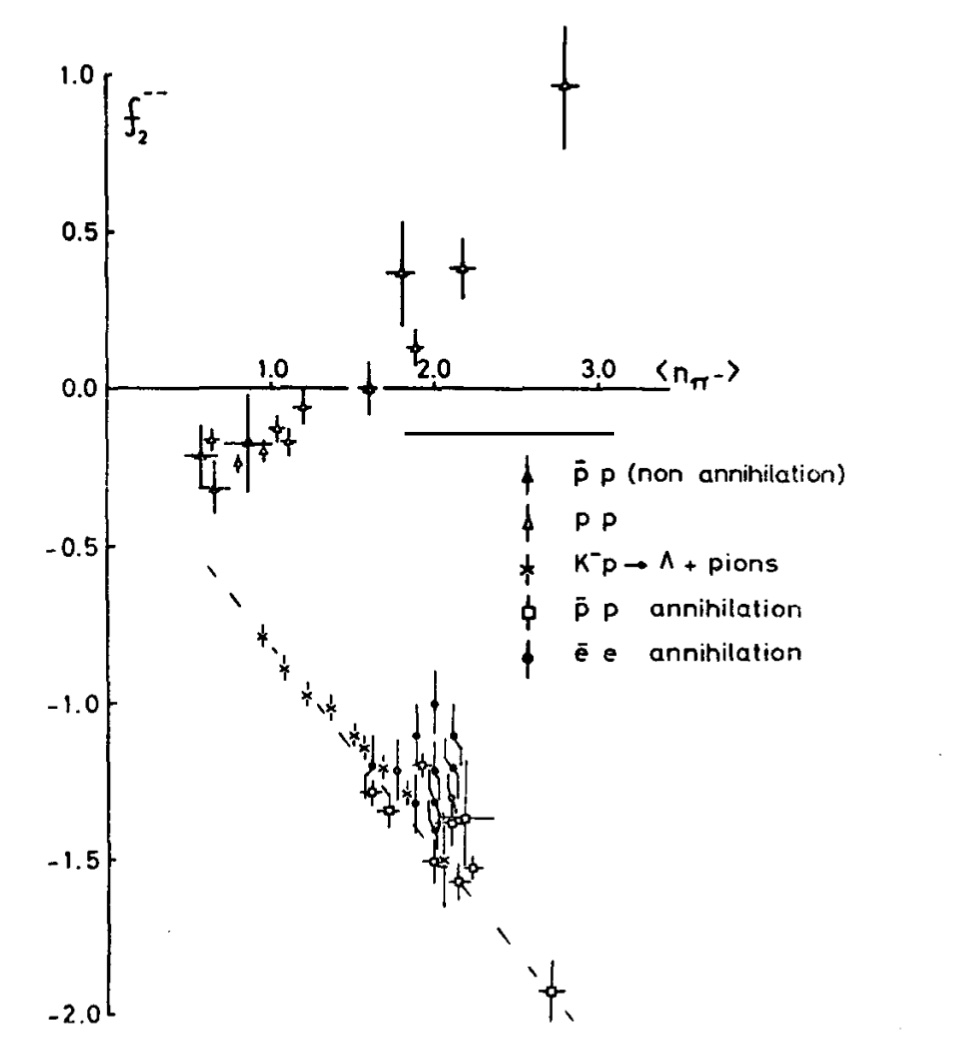}
	\includegraphics[angle=0, width=0.4\textwidth]{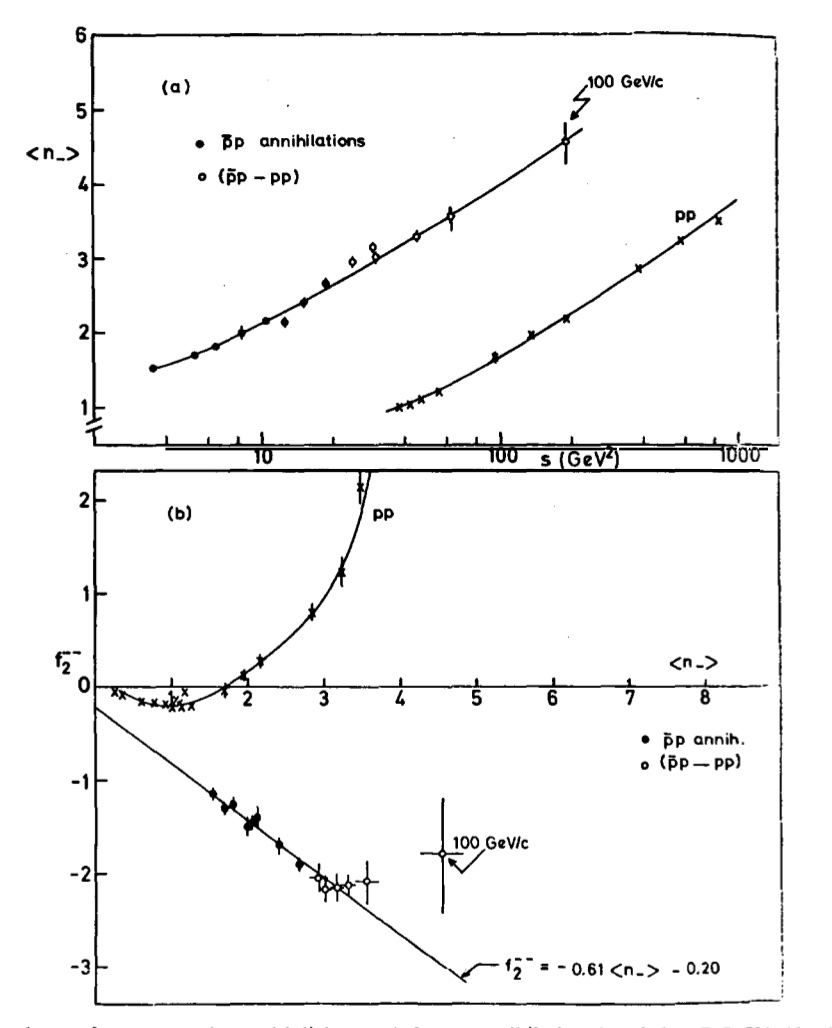}
	\caption{Experimental values of $f_2$ as a function of the mean multiplicity of negative particles $\overline n_{-}$ for annihilation and non-annihilation processes \cite{Rush}.} 
	\label{fig1}
\end{figure}

The second correlation moment calculated for the multiplicity distributions (MD) in quark and gluon jets is always positive at any energy \cite{Giov}. Considering that at low energies the $qg$-cascade is insufficiently developed and the hadronization stage predominates, we choose for it the binomial (Bernoulli) distribution with negative $f_2$ as the most suitable
\begin{equation}
	\label{eq1}
	P_n^h =  \binom{N_p}{n}\left ( \frac{\overline n_p^h}{N_p}\right )\left(1- \frac{\overline n_p^h}{N_p} \right )^{N_p-n},
\end{equation}
where index $p$ corresponds to $q$ (quark) or $g$ (gluon), $\binom{N_p}{n}$ is the binomial coefficient, $\overline n_p^h$ and $N_p$ determine the average and maximum possible number of hadrons formed from one parton $p$ at the hadronization stage and given energy.

The scheme combining two stages of MP was originally called the two-stage model 
(TSM) \cite{GDM1}. The MDs obtained in it agreed well with the data on electron-positron annihilation up to $\sim$ 200 GeV \cite{GDM1,GDM3}. Moreover, the average number of hadrons $\overline n^h$ formed from one gluon at the hadronization stage remained approximately constant and close to 1, which is in good agreement with the hypothesis of local parton-hadron duality (LoPAD) and the fragmentation nature of hadronization \cite{Muel}.

To predict the behavior of topological cross sections in proton collisions before carrying out our experiment at the SVD-2 facility located at the U-70 accelerator at IHEP, the TSM was used. It only needed to be modified for proton-proton collisions.
Including all valence quarks into $qg$-cascade showed that the hadronization gluon parameter $\overline n^h$ becomes significantly less than 1. It contradicted to $e^+e^-$-annihilation results. 

A sequential decrease of the valence quarks was leading to the growth  hadronization parameter but insignificantly. And only with their complete exclusion $\overline n^h$ did  approach 1, slightly exceeding it, in contrast to $e^+e^-$-annihilation. We explained this by the change of hadronization mechanism in the quark-gluon medium to recombination, in contrast to fragmentation in vacuum, as proposed by the head of the theoretical department of BNL B.~Muller \cite{Muel}.

Exclusion of valence quarks from the model means their conservation in the leading particles. At the same time, the birth of secondaries occurs due to energetic gluons, which we called active. That is why, TSM began was renamed the gluon dominance model (GDM) \cite{GDM4}.

This model is successfully applied to describe the MD in a three-gluon decay of heavy quarkonia \cite{Tomsk} and in proton-antiproton annihilation \cite{NPCS24}.

At present, JINR is building the NICA collider, where heavy ions, polarized protons and deuterons will be accelerated. Ion collisions will be performed at the MPD facility, and polarized particles at the SPD. Our SPD team proposes to include in its physics program the study of MP in proton collisions, started at IHEP at the SVD-2 facility, which will be able to register rare events with high multiplicity.

The phenomenological model, GDM will contribute to a more profound study of the gluon structure of the nucleon. One of the important physical problems included in the SPD program is the study of the gluon component of the proton and neutron (in the deuteron) \cite{TDR}. It should be noted that the physics program planned for the future eRHIC collider at BNL is also aimed at clarifying the role of gluons in the formation of visible baryonic matter in the Universe \cite{eRHIC}.

Experimental MD's in $e^+e^-$ annihilation have been described  in framework of TSM as soon as they became available over the course of a decade. The transition from one energy to another higher one occurs regardless of the results obtained previously. Some model parameters proved to be correlated during fitting. Therefore, it became necessary to conduct an end-to-end comparison of TSM across the entire available energy range.

In this article, the second chapter presents our results of the MD re-description in the framework of TSM (GDM) from 14 GeV up to 189 GeV. Calculations of the second correlation moment, $f_2$, by taking into account of the quark and gluon jets with following hadronization in $e^+e^-$ annihilation and explanation of the change of its sign are given in the third chapter. The fourth illustrates the behavior of TSM parameters with increasing energy in the specified interval. Some of them may indicate a change in the mechanism of MP, including the hadronization stage. The fifth chapter sums up our findings and makes predictions of average hadron multiplicity for experiments at future electron-positron accelerators (at 500 GeV and 1 TeV). 

One of the most important findings of this investigation is the study of MP is extremely sensitive to the choice of phenomenological models for its description.

\section{GDM and $e^+e^-$-annihilation}
The GDM (TSM) has been proposed to describe the MD of secondary hadrons in $e^+e^-$-annihilation. By that time, QCD already existed, and experimental MDs were partially available. In the area of applicability QCD (PT), differential-difference equations for generating functions (GF) were constructed, corresponding to the MD in quark and gluon jets \cite{KUV,Giov}.

By definition, the generation function (GF) $Q(s,z)$ for MD $P_n$ is a convolution of it with an auxiliary variable $z$
\begin{equation}
	\label{eq3}
	Q(s,z) = \sum _n {P_n z^n}, 
\end{equation}
where $s$ is the square of initial energy. $P_n(s)$ can be obtained from $Q(s,z)$ by derivation on the variable $z$
\begin{equation}
	\label{eq4}
	P_n = \frac{1}{n!}\frac{\partial ^n}{\partial z^n}Q(s,z)|_{z=0}.
\end{equation}
It is also possible to calculate the correlation moments $F_k(s)$, which are in demand when studying multiple processes (MP)
\begin{equation}
	\label{eq5}
	F_k(s) = \overline {n(n-1)(n-2)\dots (n-k+1)} = \frac{\partial ^k}{\partial z^k}Q(s,z)|_{z=1} 
\end{equation} 
(the top horizontal line means averaging over the number of particles).
The average multiplicity $\overline n$, the variance $D_2$ and the second correlation moment $f_2$ are determined through the GF:
\begin{equation}
	\label{eq6}
	\overline n = \sum _n n P_n = \frac {\partial }{\partial z}Q(s,z)|_{z=1},
\end{equation}
\begin{equation}
	\label{eq7}
	f_2 = \overline {n(n-1)} -{\overline n^2} = Q^{''}(s,z)|_{z=1} -( Q^{'}(s,z)|_{z=1})^2, D_2 = f_2 +\overline n.
\end{equation}

At the description of MP as a Markov branching process \cite{Giov}, the energy variable $Y$ \cite{Giov} is taken as an evolutionary parameter. Two elementary events are taken into account: q-bremsstrahlung ($q\to q+g$) and gluon fission ($g\to g+g$). The probabilities of these events are determined in QCD (PT). The formation of a $q\overline q$-pair from a gluon at the $qg$-cascade stage is suppressed according to QCD estimates, so it is neglected.

Differential-difference equations for the quark and gluon jets allows us to determine MD's of gluons $P_m^P$ ($P = q, g$) \cite{Giov} in them. For the quark jet, this is the Polya-Egenberger or negative binomial distribution (NBD).
\begin{equation}
	\label{eq8}
	P_m^q =\frac{k_p(k_p+1)\dots (k_p+m-1)}{m!}\left ( \frac {\overline m}{\overline m +k_p}\right )^m \left (\frac{k_p}{\overline m +k_p}\right )^{k_p},
\end{equation}
where the parameter $k_p$ is determined by the ratio of probabilities of two elementary events (quark bremsstrahlung to gluon fission), $\overline m$ is the average multiplicity of gluons. The probability of their absence ($m=0$) in a quark jet is equal to $P_0 = \left (\frac {k_p}{\overline m +k_p}\right )^{-k_p}$.
GF for MD in a quark jet (\ref{eq8}) has the following view
\begin{equation}
	\label{eq18}
	Q^{q} =\left (\frac{k_p}{k_p + \overline m}\right )^{k_p} \left[1-\frac {\overline m}{\overline m +k_p}z\right ] ^{-k_p}.
\end{equation}

For gluon jets, MD is the Farry distribution \cite{Giov}
\begin{equation}
	\label{eq9}
	P_m^g =\frac{1}{\overline m}\left ( 1-\frac{1}{\overline m}\right )^{m-1}
\end{equation}
with GF 
$$
	G(z) = \frac{z}{\overline m} \left [ 1-z\left (1-\frac{1}{\overline m}\right )\right]^{-1}.
$$
		
Energy of partons at their fission decreases up to a certain value. And then, the hadronization starts. At this stage, gluon can also continue to make fission (in the non PT QCD already) and some of them decay into $q\overline q$-pairs (we call these gluons active) forming secondary hadrons.

The TSM (GDM) can be applied to describe the MD in $e^+e^-$-annihilation. This process can be represented as the following sequence of events: the formation of a virtual photon or $Z^0$-boson, the birth of a $q\overline q$-pair, the development of a $qg$-cascade, and hadronization described by the MD (\ref{eq1}).

It is assumed that at the second stage there is no significant momentum transfer between partons, the so-called soft discolorating, which allowed the convolution of two stages to be performed and to obtain the hadron MD used to describe the \cite{GDM1,TSM} data.

According to the binomial law (\ref{eq1}) chosen on the basis of the experimental data \cite{Rush}, 
the probabilities of a single hadron being creating from a parton (a quark or a gluon) is equal to $\overline n ^h_P/ N_P$ ($P=q,g$)
We suggest that these probabilities are comparable. So, we can introduce additional parameter $\alpha $ and express the gluon  parameters of hadronization through the quark ones
$ \overline n^h_g = \alpha \overline n^h$ and $N_g = \alpha N$,
where $N = N_q$ (remove the index $q$).

The combination of two stages can be written in the following view
\begin{equation}
\label{eq19}
P_m(s) = \sum _{m=0}^{M_g} P_m^q \binom{(2+\alpha m)N}{n}
\left (\frac{\overline n ^h}{N}\right ) ^n \left ( 1-\frac{\overline n^h}{N}\right )^{(2+\alpha m)N-n},
\end{equation}
where we are limited in summation (\ref{eq19}) to a finite number of active gluons ($M_g$).

To compare (\ref{eq19}) with experimental data we add the normalization factor $\Omega $ before summation. This expression includes two parameters to describe the $qg$-cascade ($\overline m$ and $k_p$) and three ($\overline n^h$, $N$ and $\alpha $) for the  hadronization stage. In this study, within the framework of GDM, we present a re-description of experimental data on $e^+e^-$ annihilation by the expression (\ref{eq19}). We use improved modern software and existing experimental data in a wide energy range ($\sim $  10 $\div $ 200 GeV). The behavior of these parameters with increasing energy provides us additional information about the $qg$-cascade and hadronization. 

\begin{figure}[H] 
	\leavevmode
	\centering
	\includegraphics[angle=0, width=0.45\textwidth]{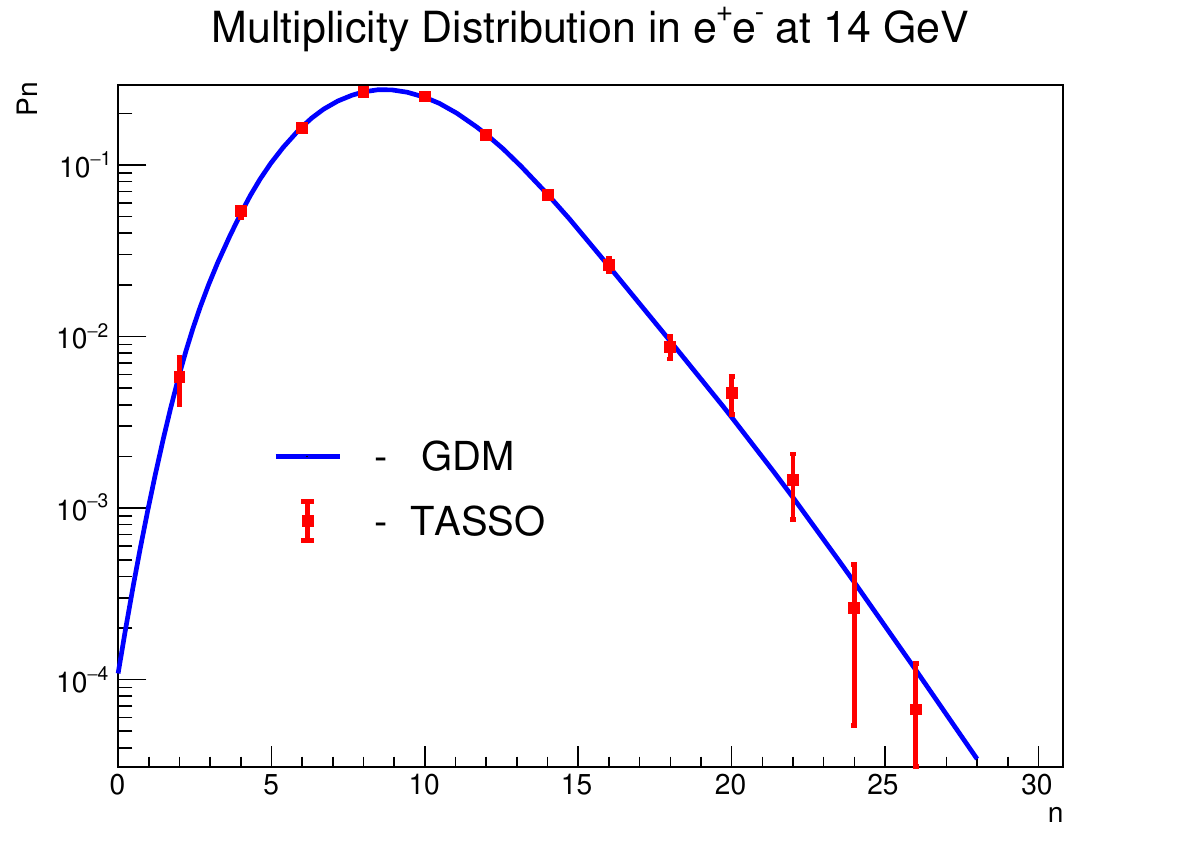}
	\includegraphics[angle=0, width=0.45\textwidth]{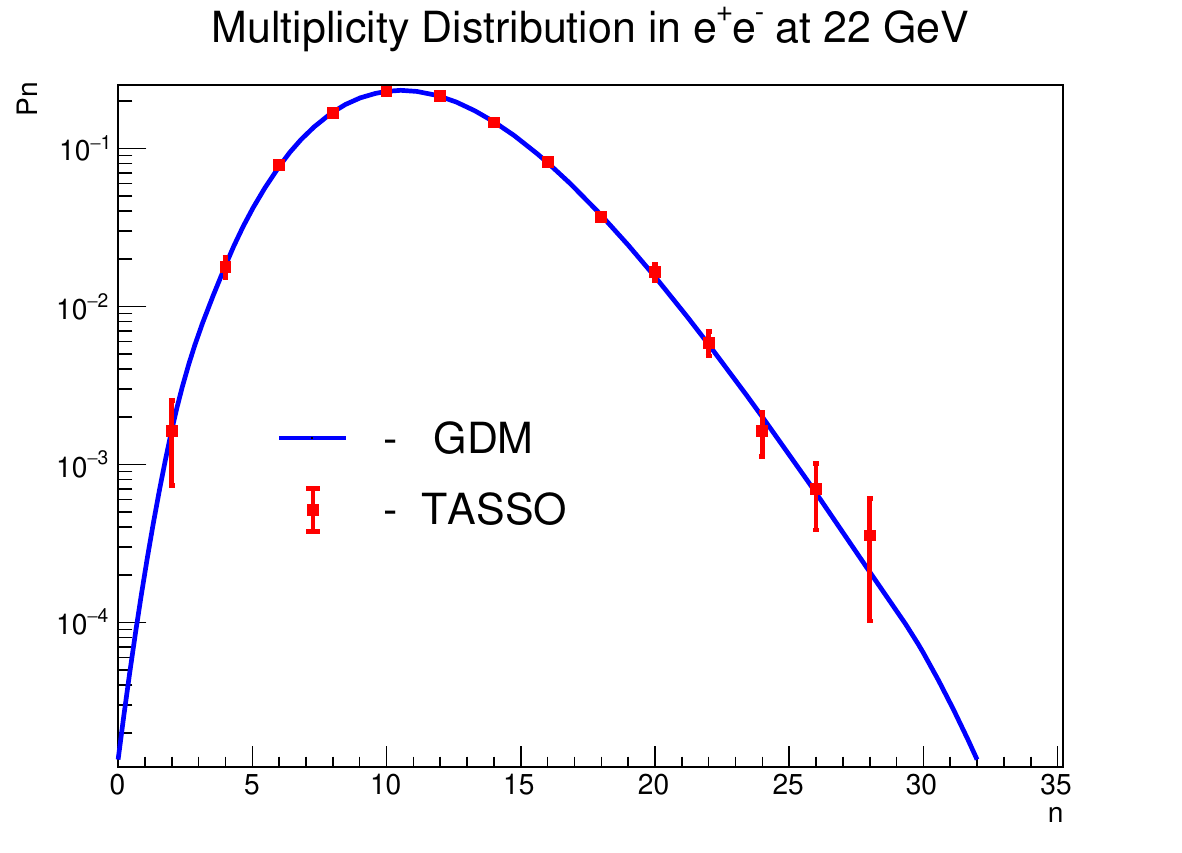}
    \includegraphics[angle=0, width=0.45\textwidth]{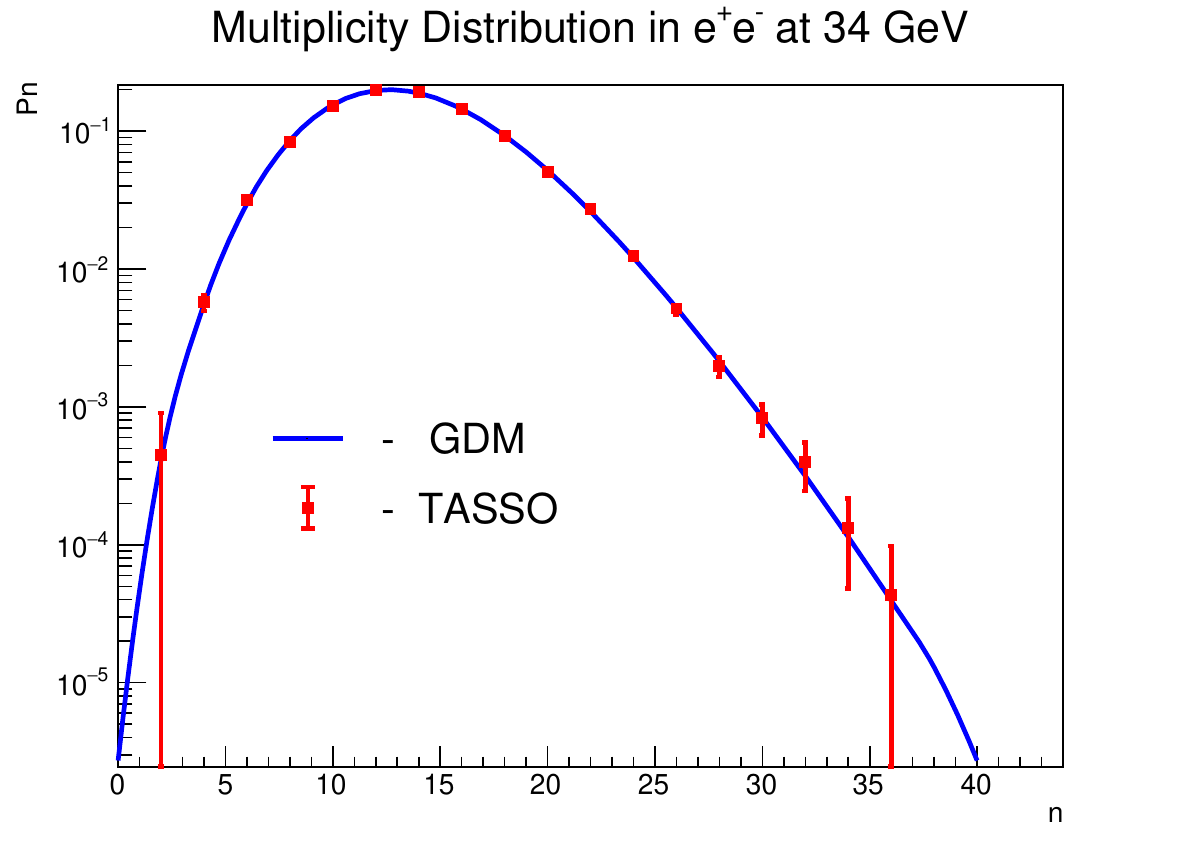}	
    \includegraphics[angle=0, width=0.45\textwidth]{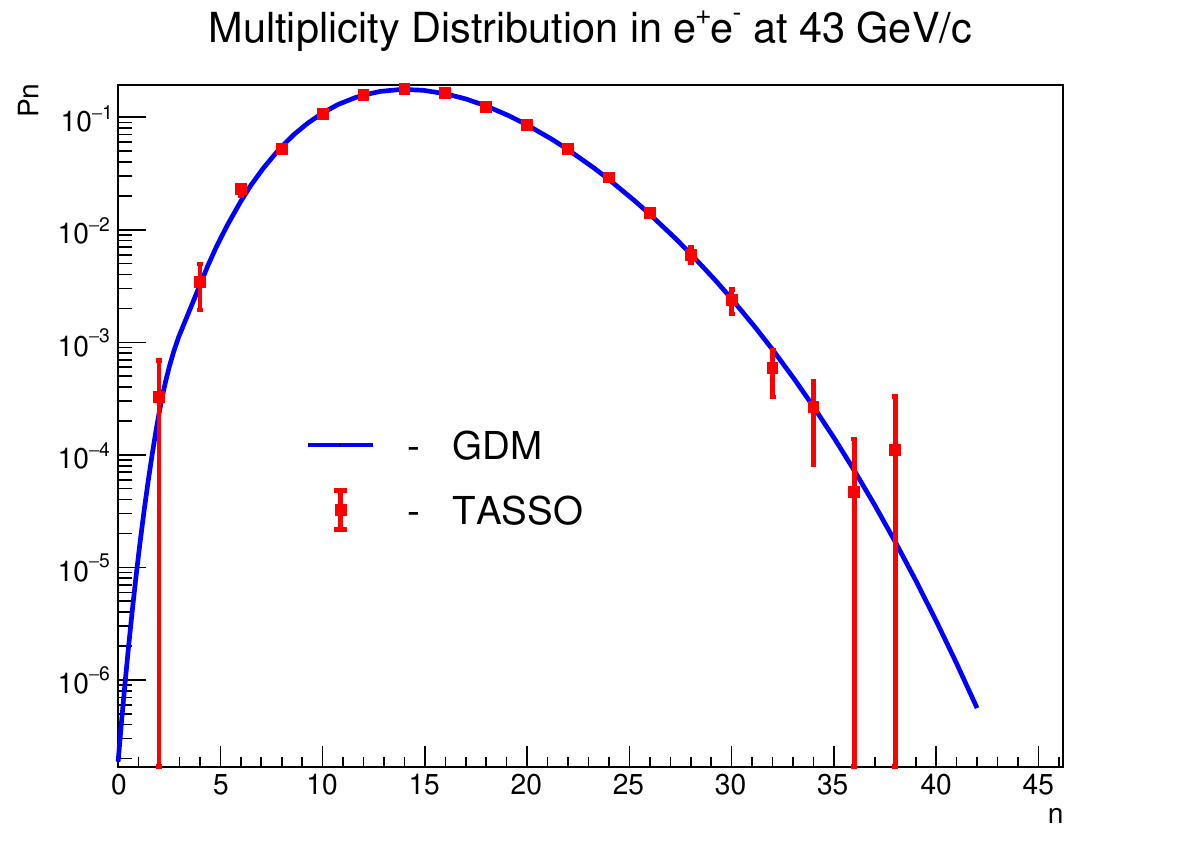}
	\caption{The description of multiplicity distributions, $P_n$, in $e^+e^-$ annihilation at 14, 22, 34 and 43 GeV \cite{ee43} (red boxes) into framework of GDM (blue line)} 
	\label{fig2}
\end{figure}

\begin{figure}[H] 
	\leavevmode
	\centering
	\includegraphics[angle=0, width=0.45\textwidth]{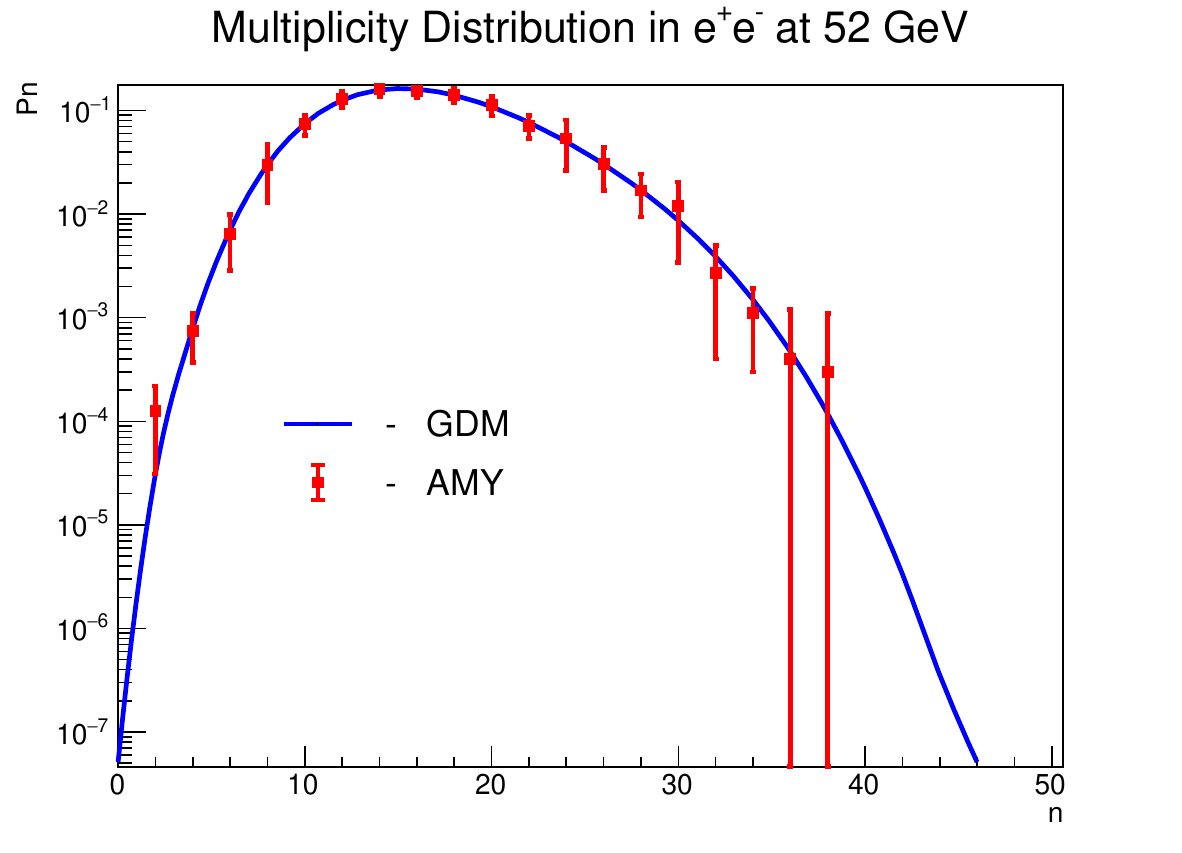}
	\includegraphics[angle=0, width=0.45\textwidth]{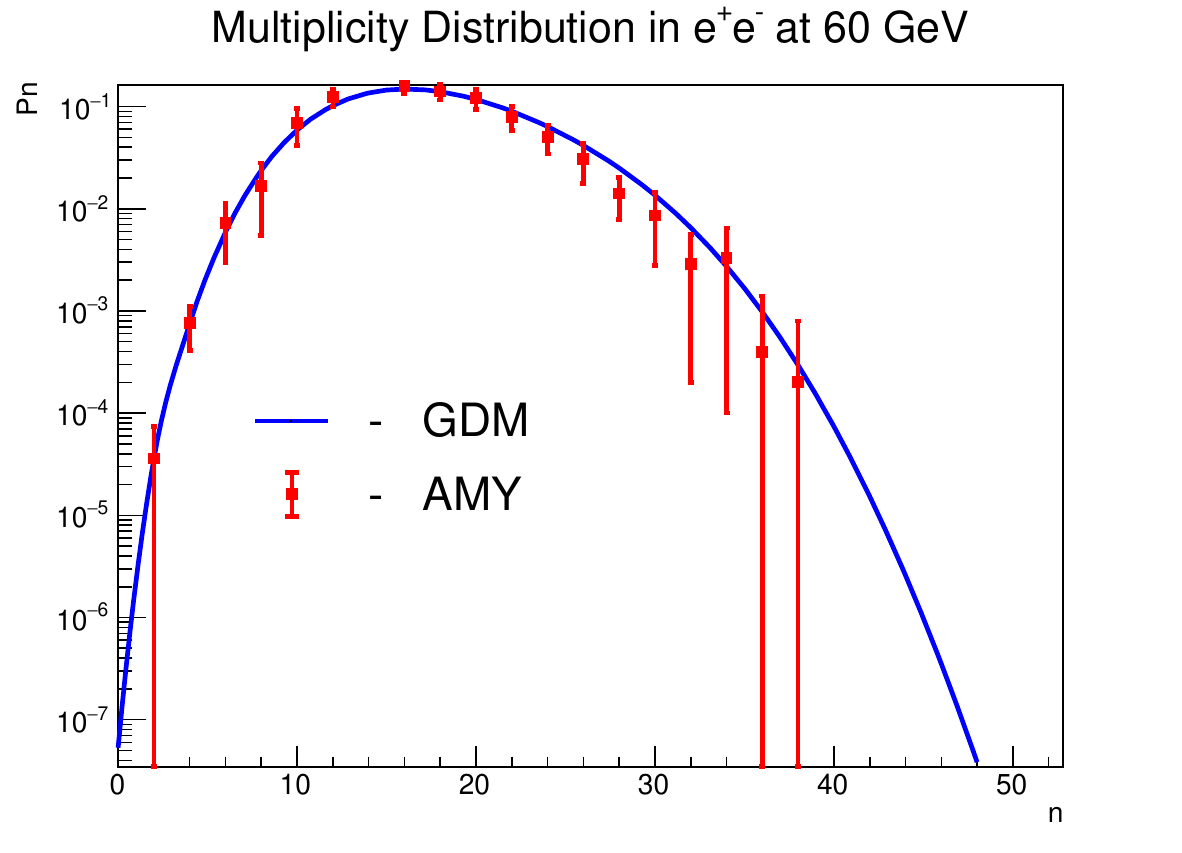}
	\caption{The description of multiplicity distributions, $P_n$,  in $e^+e^-$ annihilation at 52 and 60 GeV \cite{ee50} (red boxes) into framework of GDM (blue line).} 
	\label{fig4}
\end{figure}

Comparison of the GDM with data of TASSO Collaboration \cite{ee43} at 14, 22, 34 and 43 GeV are presented in Figures \ref{fig2}. AMY Collaboration \cite{ee50} carried out
measurements of MD in a more detailed region. It performed eight measurements from 50 to 61.4 GeV with intervals of 1--2 GeV between them. Figure \ref{fig4} illustrates the description of MD by GDM at 52 and 60 GeV. Note that the remaining six distributions are also consistent with the data well.

\begin{figure}[H] 
	\leavevmode
	\centering
	\includegraphics[angle=0, width=0.45\textwidth]{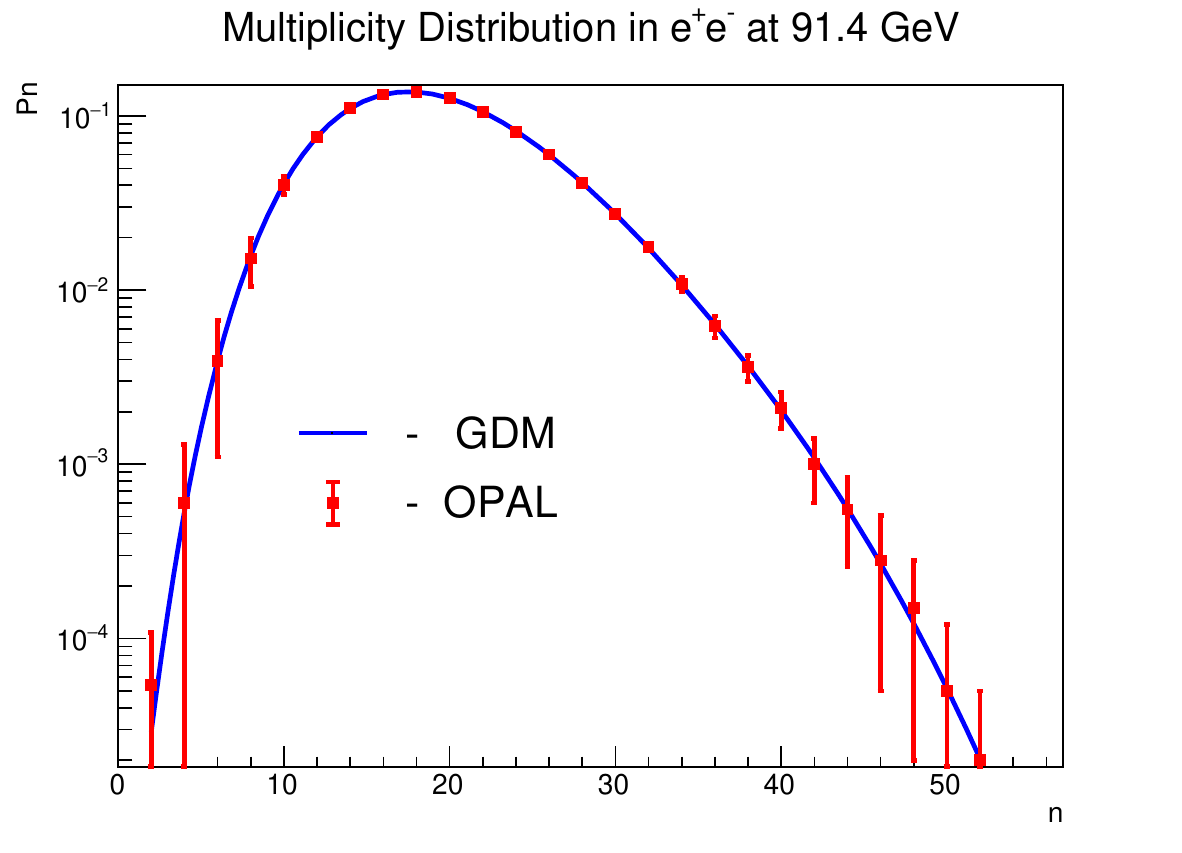}
	\includegraphics[angle=0, width=0.45\textwidth]{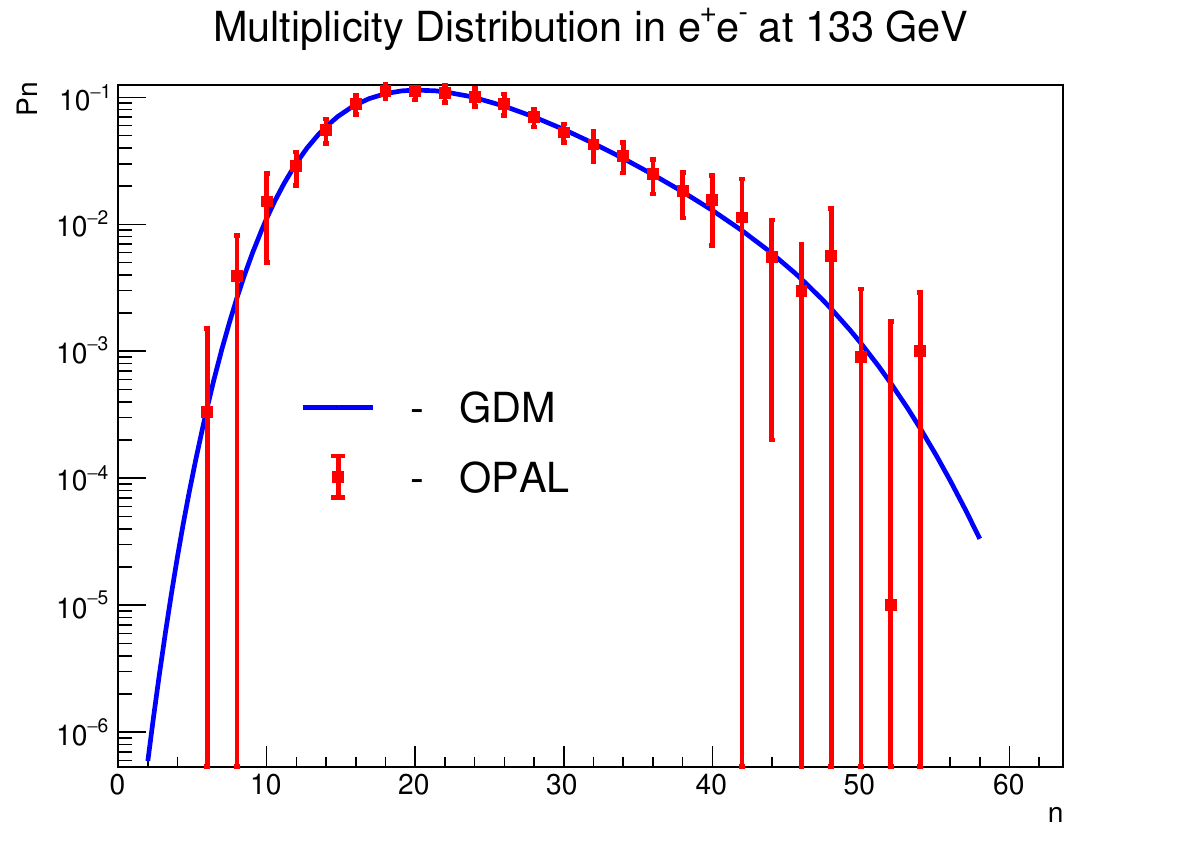}
	\hfill{}
	\hfill{}
	\includegraphics[angle=0, width=0.45\textwidth]{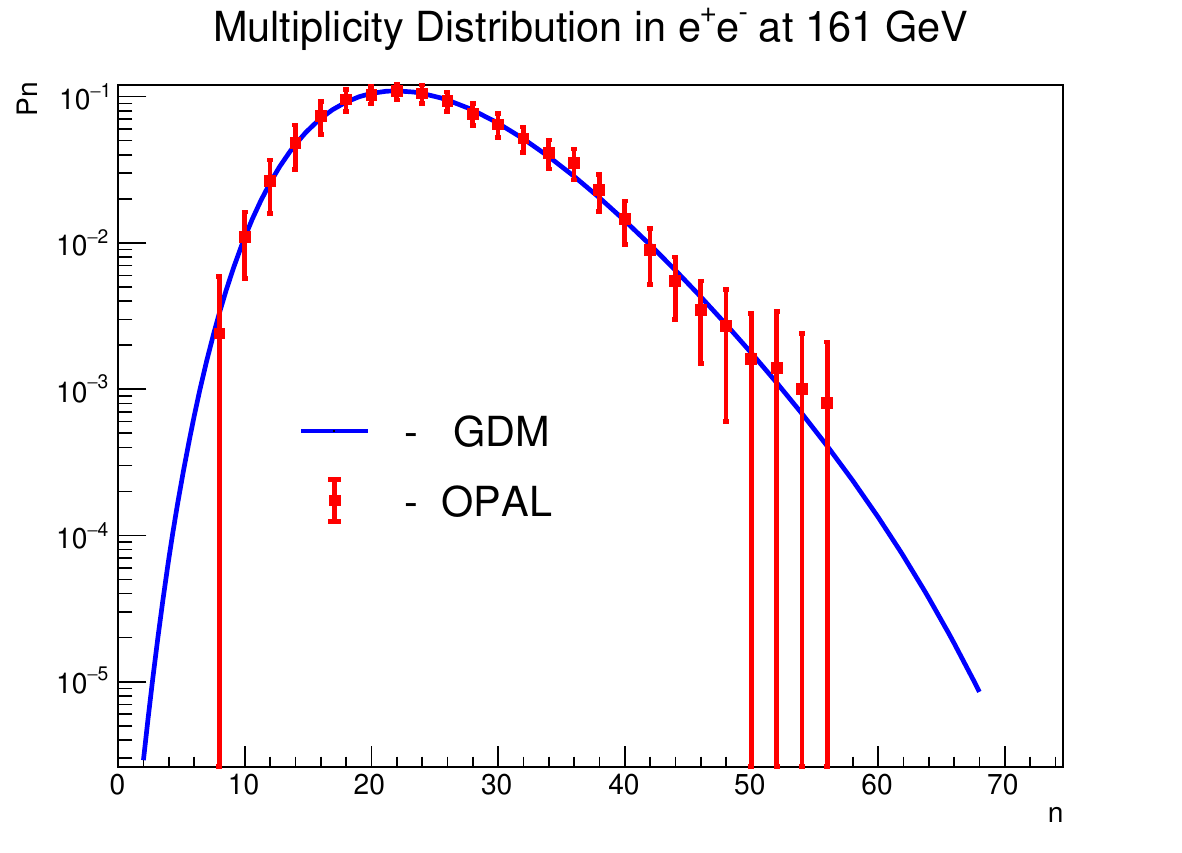}
    \includegraphics[angle=0, width=0.45\textwidth]{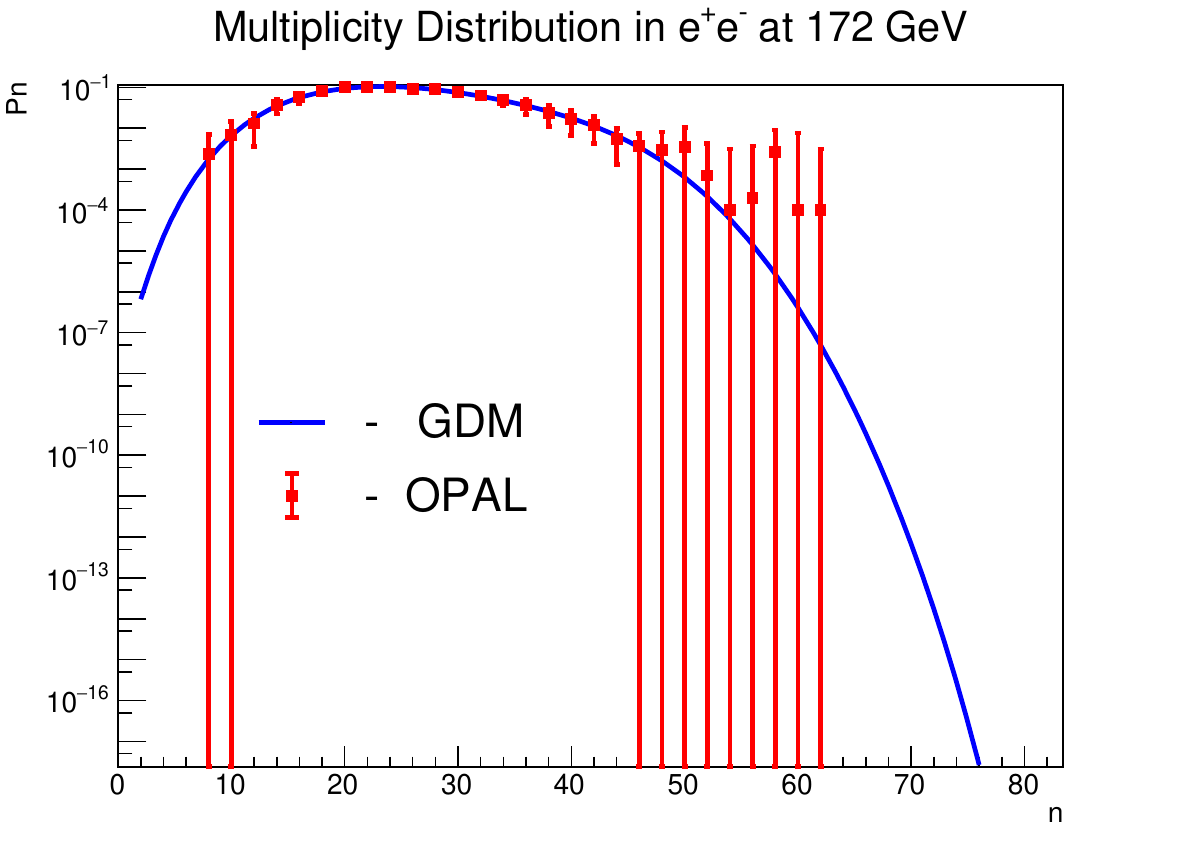}
    \hfill{}\hfill{}
    \includegraphics[angle=0, width=0.45\textwidth]{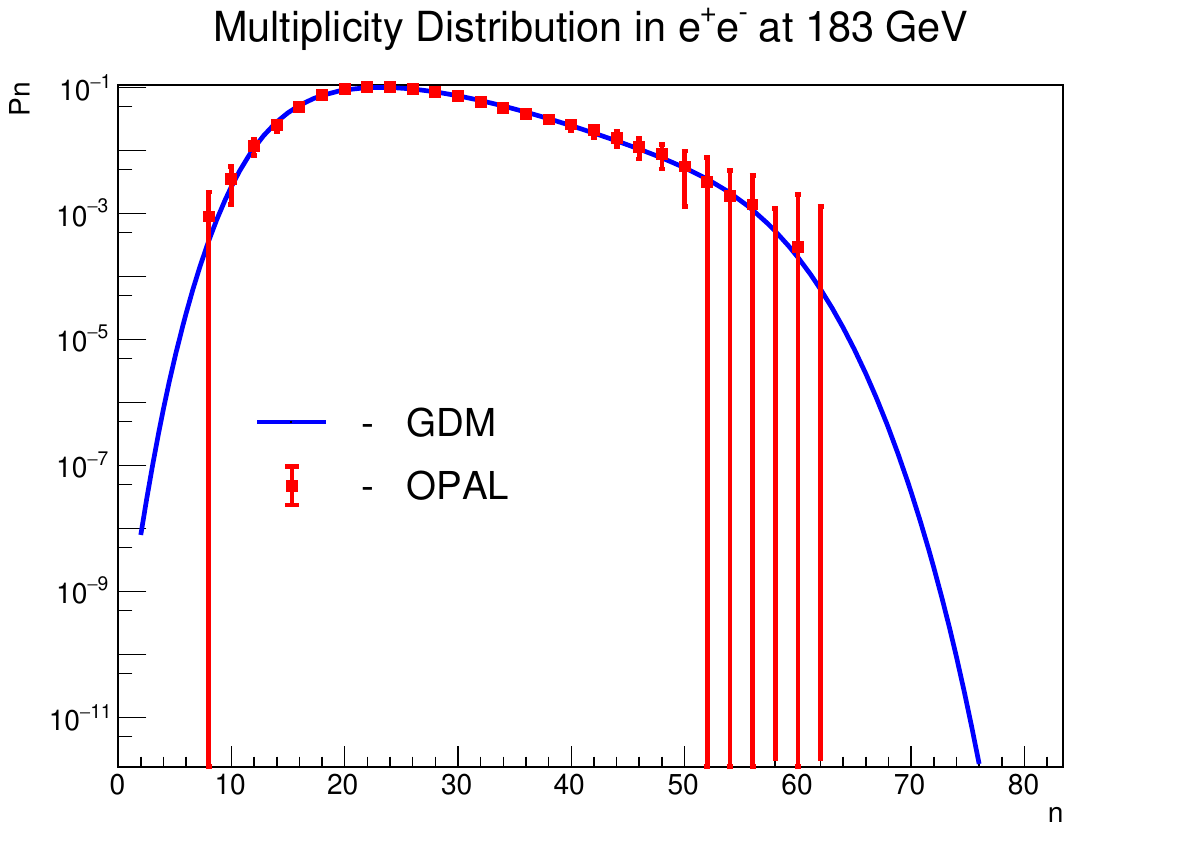}
    \includegraphics[angle=0, width=0.45\textwidth]{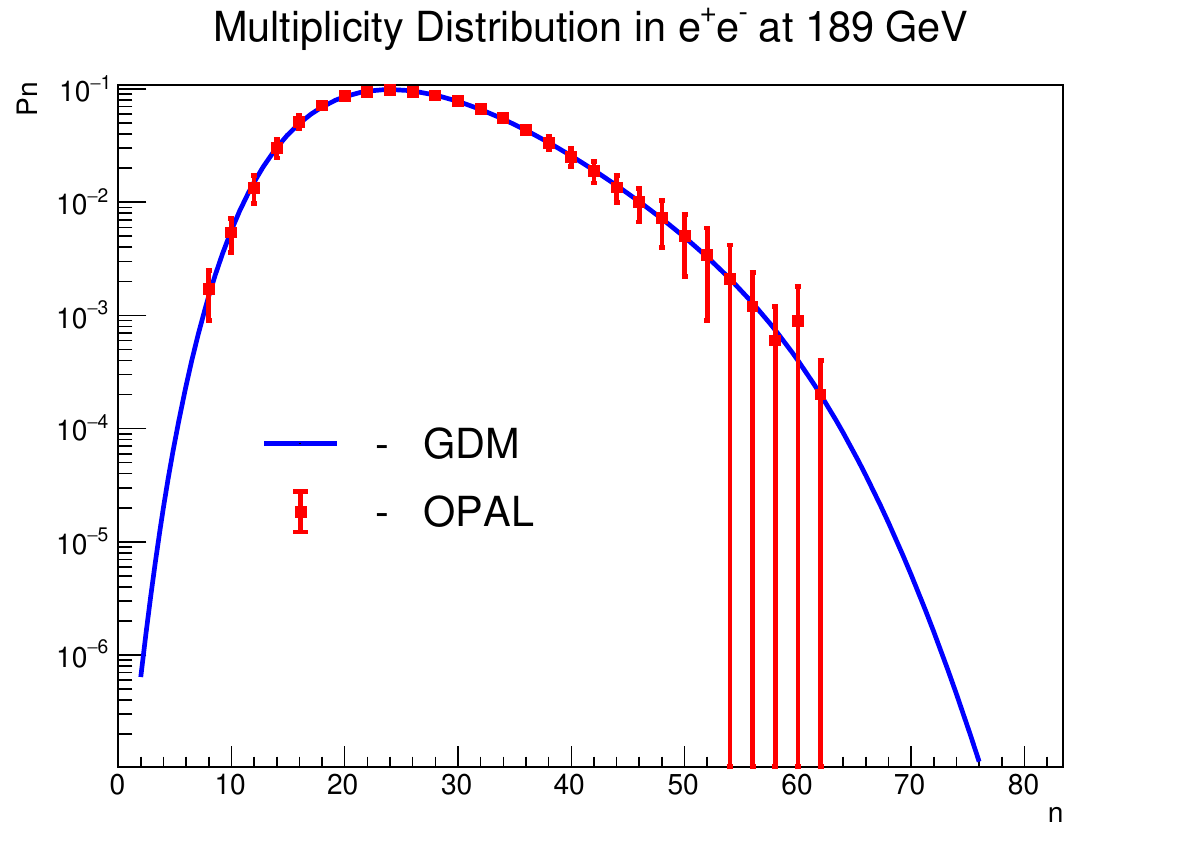}	
	\caption{The description of multiplicity distributions, $P_n$,  in $e^+e^-$ annihilation at 91.4 near of the $Z^0$-boson creation \cite{eeZ}, 133 \cite{ee131}, 161 \cite{ee161}, 172, 183 and 189 GeV \cite{ee180} (red boxes) into framework of GDM (blue line).} 
	\label{fig5}
\end{figure}

MD of the remaining  available data at higher energies were measured at four setups (OPAL, DELPHI, SLD and ALEPH). We made our choice on data of OPAL Collaboration  \cite{ee180}. In Figure \ref{fig5}  we present MD (top, left) near the $Z^0$-boson
creation at 91.4 GeV) \cite{eeZ}. The description of OPAL data at 133 GeV (top, right) \cite{ee131} by GDM and at 161 GeV (bottom, left) \cite{ee161} confirm good description of them. And at least, in Figure \ref{fig5} you can see the agreement of data \cite{ee180} with the expression ((\ref{eq19}) at energies 172, 183 and 189 GeV.

In all these Figures data are shown by red boxes, and blue line is used for the model line. Values of $\chi ^2$  for all cases do not exceed a few unites. Particularly good agreement is achieved in the area of high multiplicity (HM) 
(we often call it the high-multiplicity tail of the distribution). It is known that this region is often cannot be described by the existing Monte Carlo events generators. Our scheme of the unification excludes this trouble.

\section{Second correlative moment}
The GF for MD (\ref{eq19}) has the view
\begin{equation}
	\label{eq20}
	Q(s,z) = \sum _{m=0}^{M_g} P_m^q 
	\left [1+ \frac{\overline n ^h}{N}(z-1) \right ]^{(2+\alpha m)N},
\end{equation}
where $P_m^q$ is the gluon distribution in a q quark jet (\ref{eq8}).
The average multiplicity of hadrons in $e^+e^-$ annihilation is defined as the first derivative of the generating function with respect to the auxiliary variable $z$ 
\begin{equation}
	\label{eq21}
\overline n(s) = \frac{\partial Q(z,z)}{\partial z} \left |_{z=1}=	(2+\alpha \overline{m})\bar n^h. \right .
\end{equation}	
	
The second correlation moment, calculated from the convolution of two stages can obtained from (\ref{eq20}) also
\begin{equation}
	\label{eq22}	
	f_2 = F_2 - F_1^2 = \left[\alpha ^2 \frac {\overline m ^2}{k_p}+\alpha ^2 \overline m - \frac {2+\alpha \overline m}{N}\right ](\overline n^h)^2,
\end{equation} 
where $F_1$ = $Q(s,z)^{'}|_{z=1}$ = $\overline n(s)$, $F_2$ = $Q(s,z)^{''}|_{z=1}$ = $\overline {n(s)(n(s)-1)}$. 
It takes negative values at low average gluon multiplicity $\overline m$ and becomes positive as it increases with energy. The parameter $\alpha $ was introduced  as the ratio of hadronization parameters ($\overline n^h_g/\overline n^h_q$). GDM gives values for $\alpha $ is less than 1 and $k_p$ is greater > 3. At low gluon multiplicity ($\overline m$), the main contribution to the expression (\ref{eq22}) is made by the negative term, the first two terms can be neglected.

The second correlative moments, $f_2$, estimate from the model parameters is consistent with the experimental data. At 14 GeV, it is close to zero in the limits of errors. Then it becomes positive and continue to grow. Negative values of $f_2$ belong to lower energies. In high energy physics, it is also useful to include into analysis higher-order moments, not only factorial but also cumulative ones, as well as their ratios, for studying MP. Previous studies of them confirmed the change (growth) in oscillation periods as a function of their order at energies above the $Z^0$-boson formation threshold \cite{Mom}.

\section{GDM parameters}
After describing of MD into framework of GDM, it would be interesting to analyze the energy behavior of the model parameters. It should be noted that, although the number of parameters  is equal only to six (including a normalization factor close to 2), the function (\ref{eq19}) is quite complex and nonlinear. Furthermore, we know very little about their true values. In addition, our approach was simplified by the assumption about comparable probabilities for a quark and gluon to create single hadron through passing the hadronization stage. 

As the number of gluons in the cascade more than two quarks (more than 20) and increases with energy (up to 50), and the accuracy of fitting the quark parameters will be lower than for gluon ones. We obtain their values by the GDM  description of data. Although it is not yet possible to theoretically calculate these parameters, the results obtained from comparison of the data, are entirely consistent with their qualitative behavior with increasing energy such as a growth of the average gluon multiplicity.

\begin{figure}[H] 
	\leavevmode
	\centering
	\includegraphics[angle=0, width=0.45\textwidth]{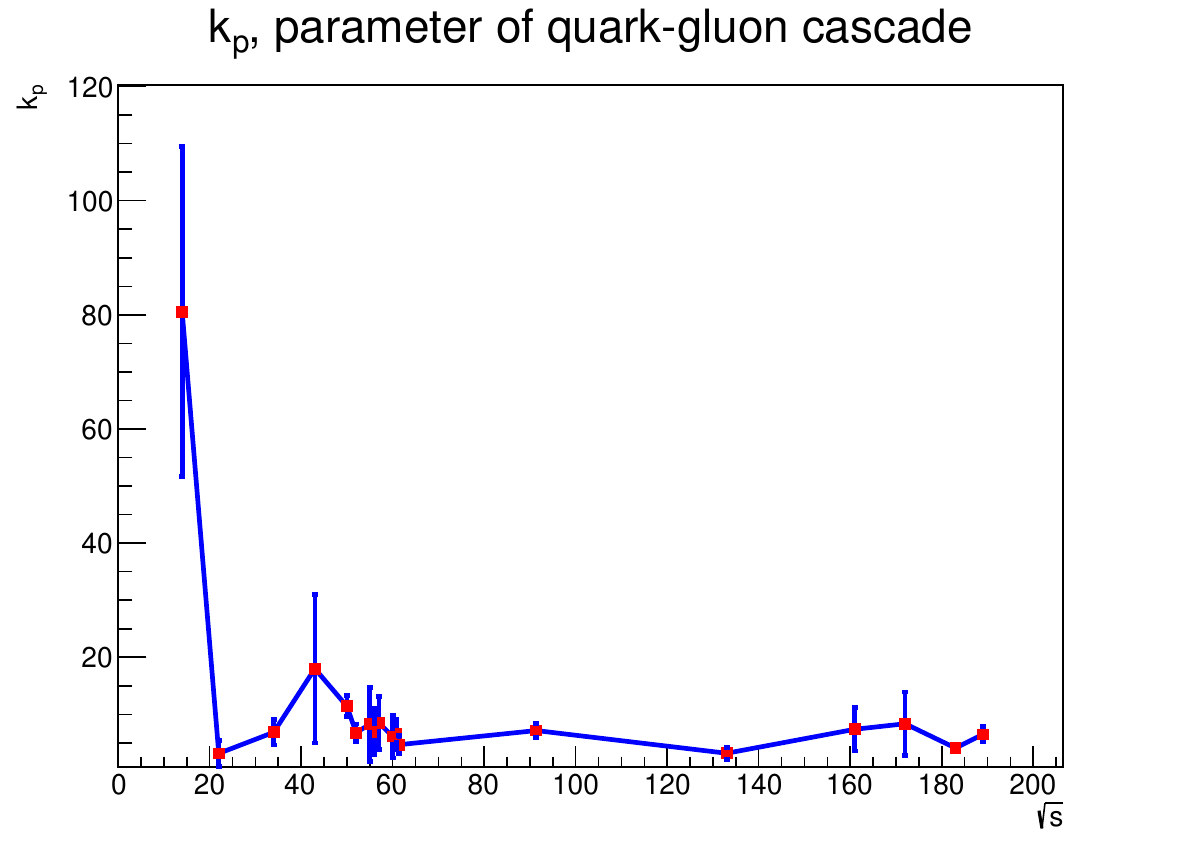}
	\includegraphics[angle=0, width=0.45\textwidth]{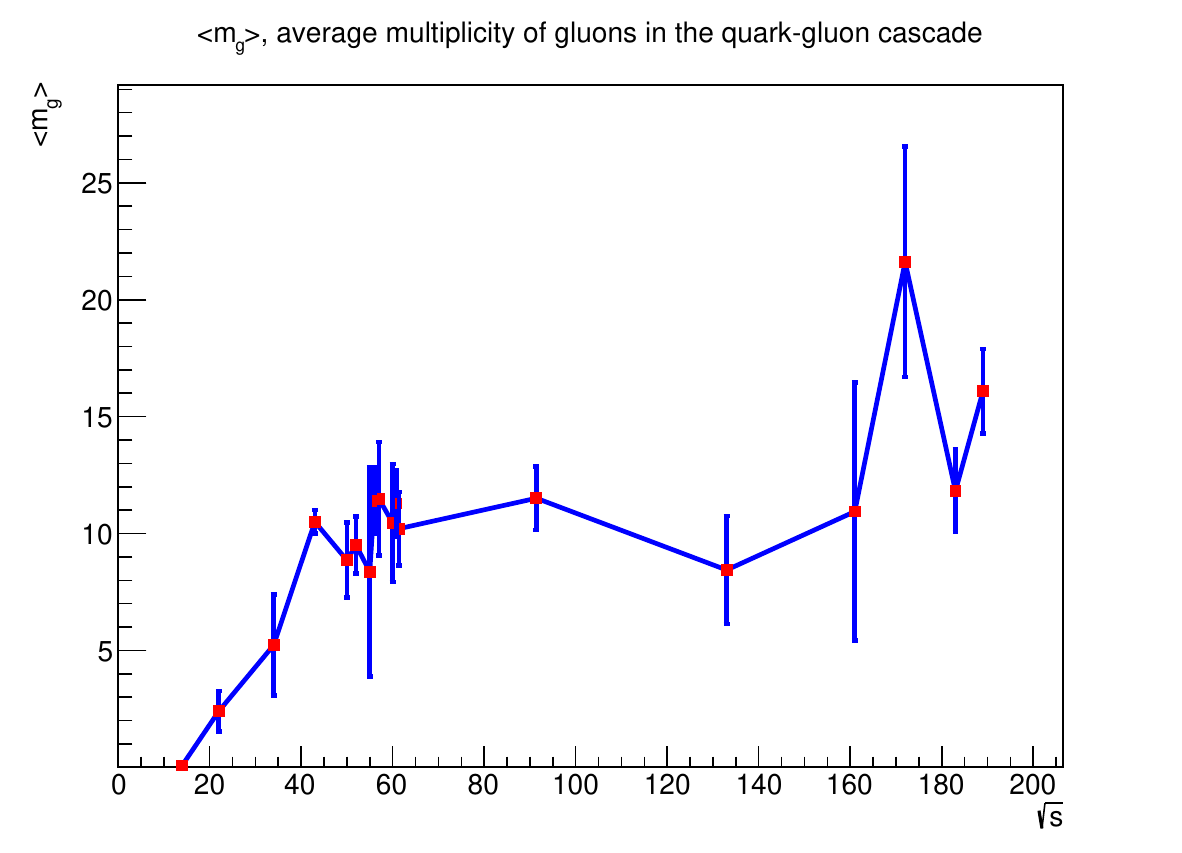}
	\caption{Parameters of GDM, $k_p$ (on the left) and $\overline m$ (on the right).} 
	\label{fig8}
\end{figure}

We begin our analysis with a quark-gluon stage.  The first parameter among them is $k_p$. It is introduced as a ratio of probability of two elementary events: a $q$-bremsstrahlung ($q \to q + g$) to a gluon fission ($g \to g +g $). It is shown in Figure \ref{fig8} (on the left).

This parameter outreaches unity across the entire range studied, and at the  minimum researched energy, 14 GeV, it exceeds 80, and then drops sharply and does not exceed 10, remaining greater than 3 and tending to decrease slightly. Thus, the contribution of gluon fission at 14 GeV is significantly suppressed, but at 22 GeV it increases sharply, by 10 times, and its fraction remains almost the same up to 200 GeV. The appearance of additional gluon fission, as it is predicted by QCD, leads to a broadening of the jets, i.e., a more developed cascade. Quark jets remain harder and narrower.  

In the energy region higher than 14 GeV $k_p$ keeps being constant and has the fitting value 5.39 $\pm $ 0.40 despite large oscillation which is stipulated by a strong correlation between parameters. It tells us about steady-state process at first stage.  The gluon fission gives contribution to all number of gluons the fifth fraction before starting of hadronization.

It should be noted that at 14 GeV, the second correlation moment is close to zero. This indicates that at lower energies, the hadronization stage is dominant over the cascade, and gluon fission are practically suppressed. In this case, $P_m^q$ in expression (\ref{eq19}) could be replaced by the Poisson distribution (the simplest flow of events) accordingly to \cite{Giov}. After final analyzing the model parameters, it will be interesting to test this assumption using the PLUTO data at 9.4 GeV \cite{PLUTO}. At this energy $f_2$ is negative (-2.4). 

The second GDM parameter, $\overline m $, denotes the average multiplicity of gluons appeared at the end of a $qg$-cascade. The Figure \ref{fig8} (right) illustrates its fast growth at low energies  (up to 60 GeV). Then it demonstrates steady and near-logarithmic growth despite fluctuations and errors. Theoretical estimations from \cite{Giov} confirms such behavior. In QCD, the average number of gluons in $qg$-cascade of $q$-jet is determined by the expression (see notation in \cite{Giov}) 
\begin{equation}
\label{eq23}
\overline m(s) = k_p (e^{AY} -1),
\end{equation}
where $k_p = \tilde A /A$, $Y \sim \log (1+\log s/s_0) $ -- an evolution parameter, $\tilde A$ and $A$ are proportional to the probabilities of $q$-bremsstrahlung and gluon fission, respectively. At low energies the mean number of gluons is proportional to $\tilde A Y$ (it depends on only q-bremsstrahlung). 

\begin{figure}[H] 
	\leavevmode
	\centering
	\includegraphics[angle=0, width=0.45\textwidth]{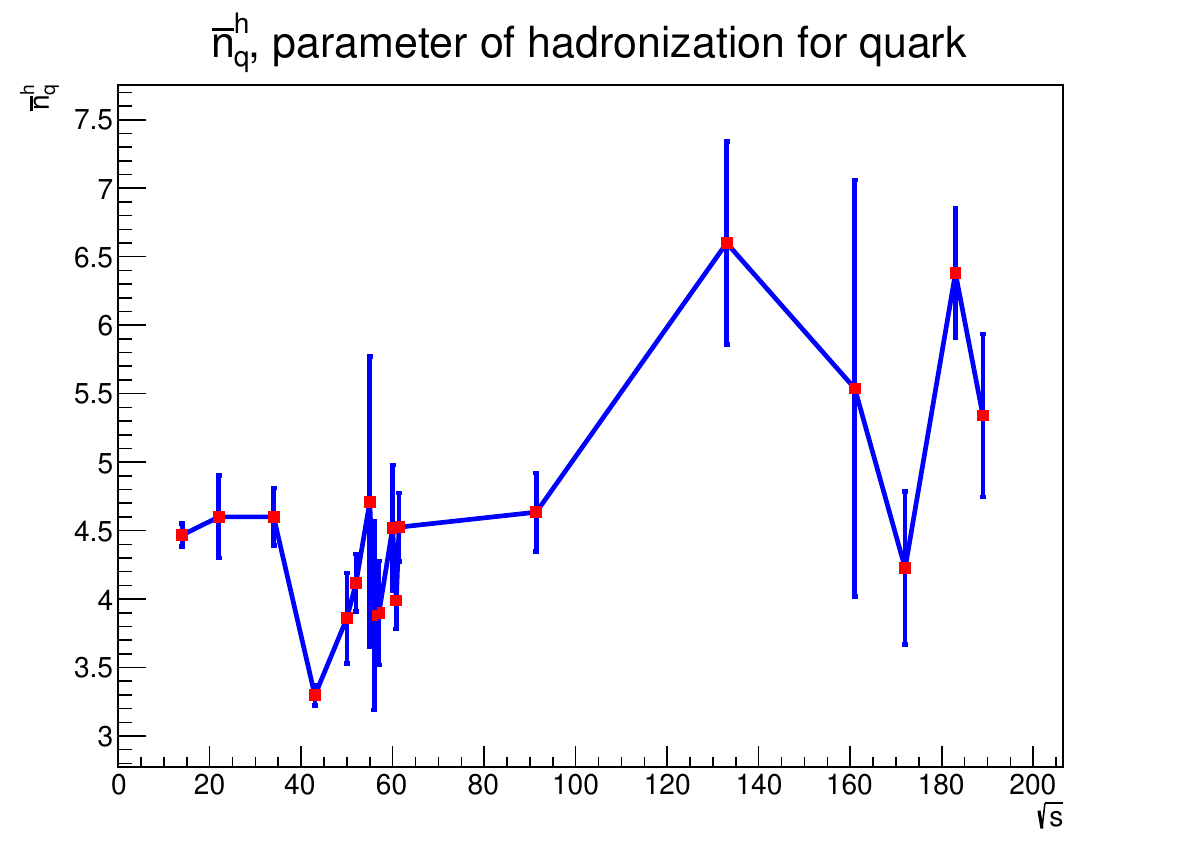}
	\includegraphics[angle=0, width=0.45\textwidth]{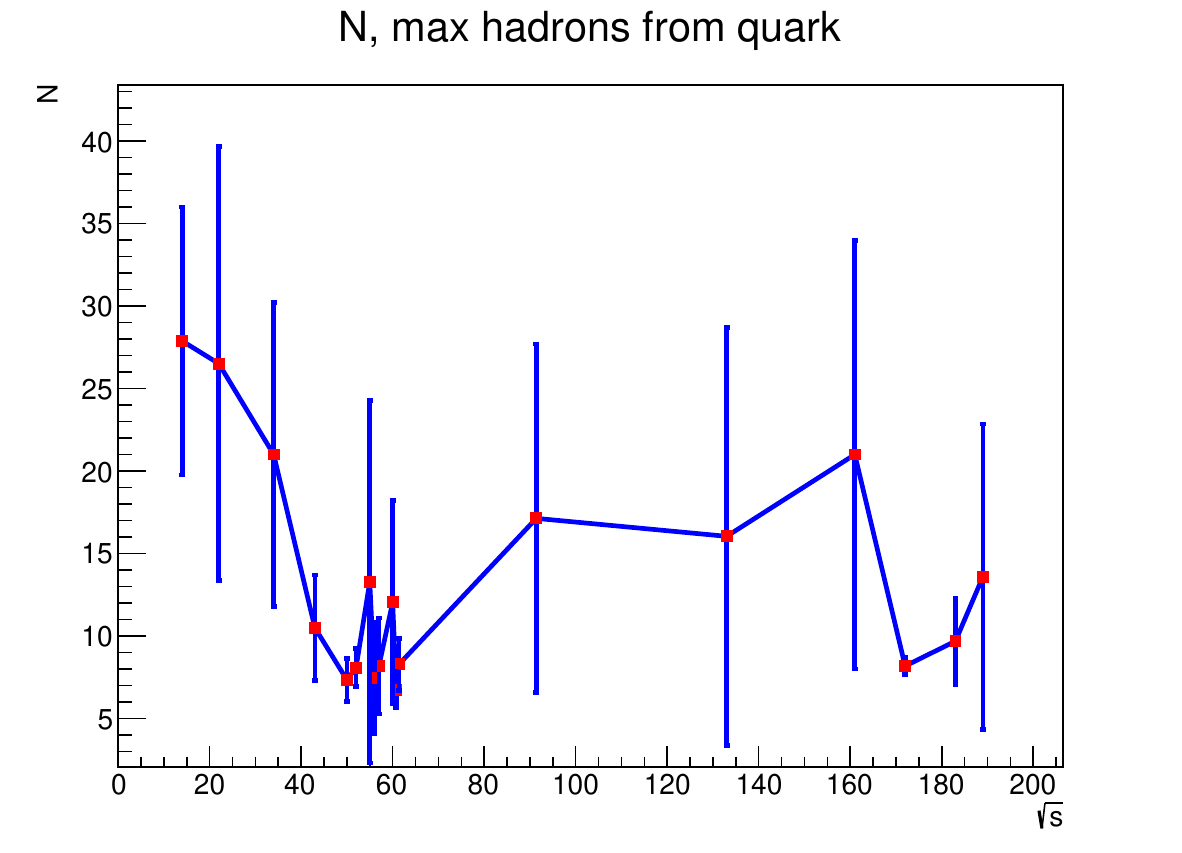}
	\caption{Parameters of hadronization for a quark jet: $\bar n^h_q$ (on the left) and $N$ (on the right).} 
	\label{fig9}
\end{figure}

\begin{figure}[H] 
	\leavevmode
	\centering
    \includegraphics[angle=0, width=0.45\textwidth]{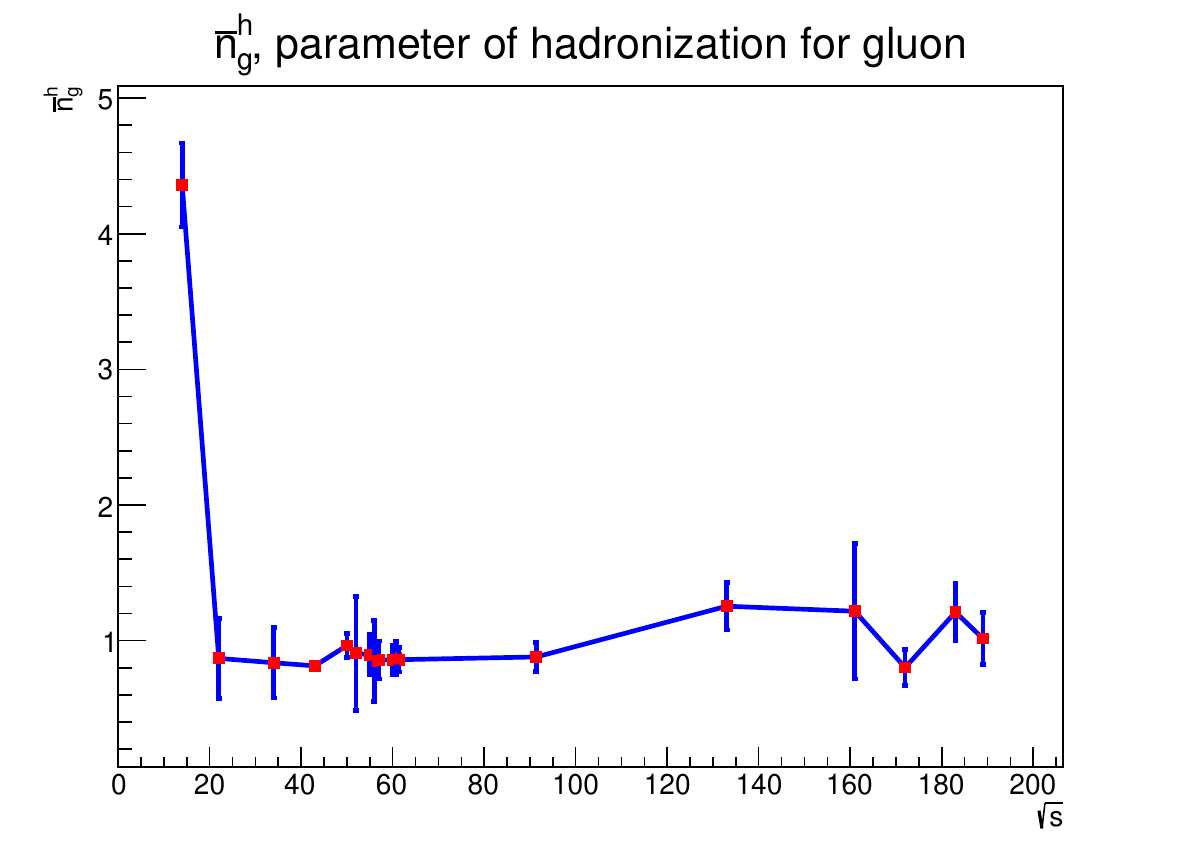}
	\includegraphics[angle=0, width=0.45\textwidth]{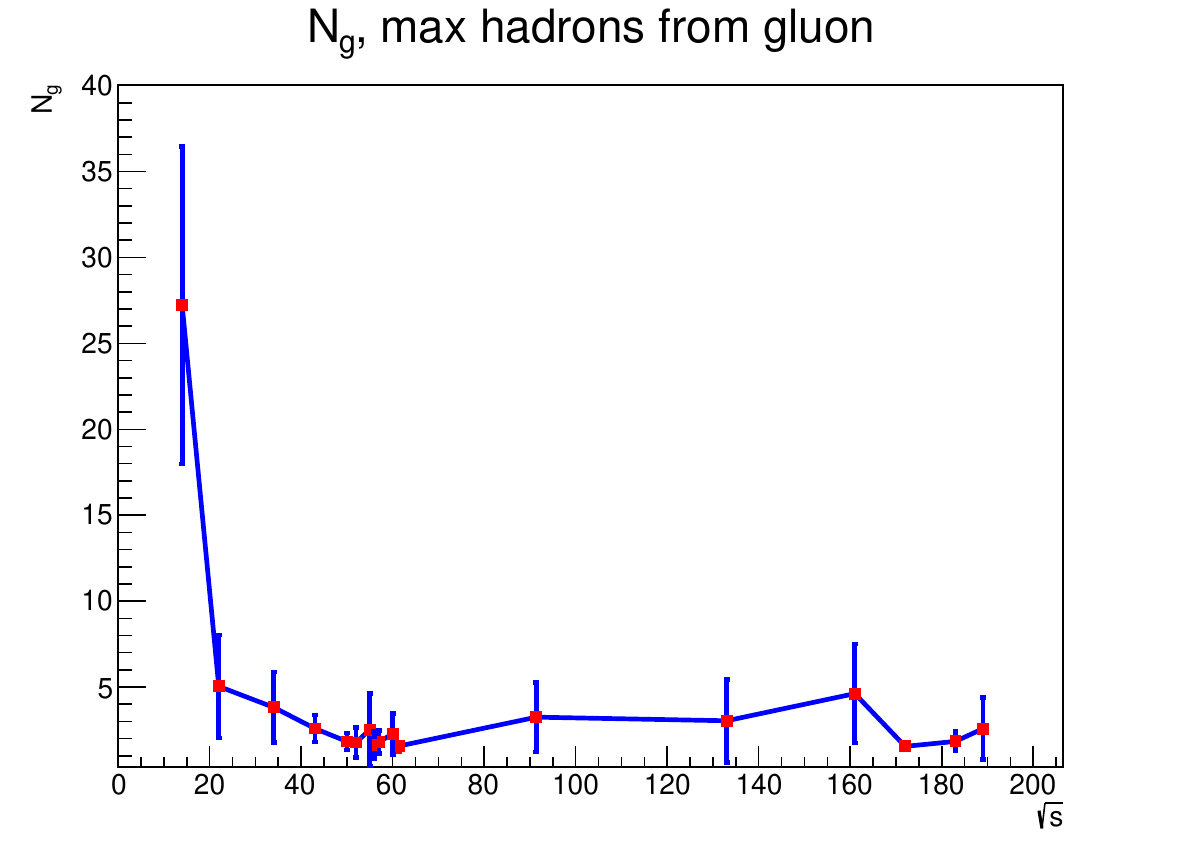}
	\caption{Parameters of hadronization for a gluon jet: $\bar n^h_g$ (on the left) and $N_g$ (on the right).} 
\label{fig11}
\end{figure}

Parameters of a $q$-jet at the hadronization stage $\overline n^h$, $N$, (in Figure \ref{fig9}) are determined in direct way from fitting, for a $g$-jet,  $\overline n^h_g$, $N_g$, are determined indirectly by using $\alpha $ (Figure \ref{fig11}). The mean number of hadrons formed from single quark at its passing hadronization is in the limits about 4--6 particles with a weak growth. The maximum number of hadron from a single quark decreases at first (part of its energy takes gluons) and then does not change.  

The opposite picture it observed for gluon parameters (Figure \ref{fig11}).
They take big values at 14 GeV, fall and remain almost constant. $N_g$ does not exceed 5. The average multiplicity of gluon, $\overline n^h_g$, at energy lower then $\sim $ 130 GeV is less then 1 ($\sim $0.9). This corresponds to the fragmentation mechanism of hadronization in vacuum, when one gluon fragments into one hadron. It is also consistent with the hypothesis of local parton-hadron duality (LoPAD), proposed for the relationship between parton and hadron average multiplicities \cite{QCD1}. 

With following increasing of energy it takes on a value that already exceeds one ($\sim $ 1.2). Such behavior we observed in hadron interactions \cite{GDM3}. Mechanism of hadronization in $pp$ collisions is recombination. In this case it occurs in dense quark-gluon medium.  From this we conclude, that such behavior of $\overline n^h_g$ indicates a change of hadronization mechanism, although not so notable.

\begin{figure}[H] 
	\leavevmode
	\centering	
    \includegraphics[angle=0, width=0.45\textwidth]{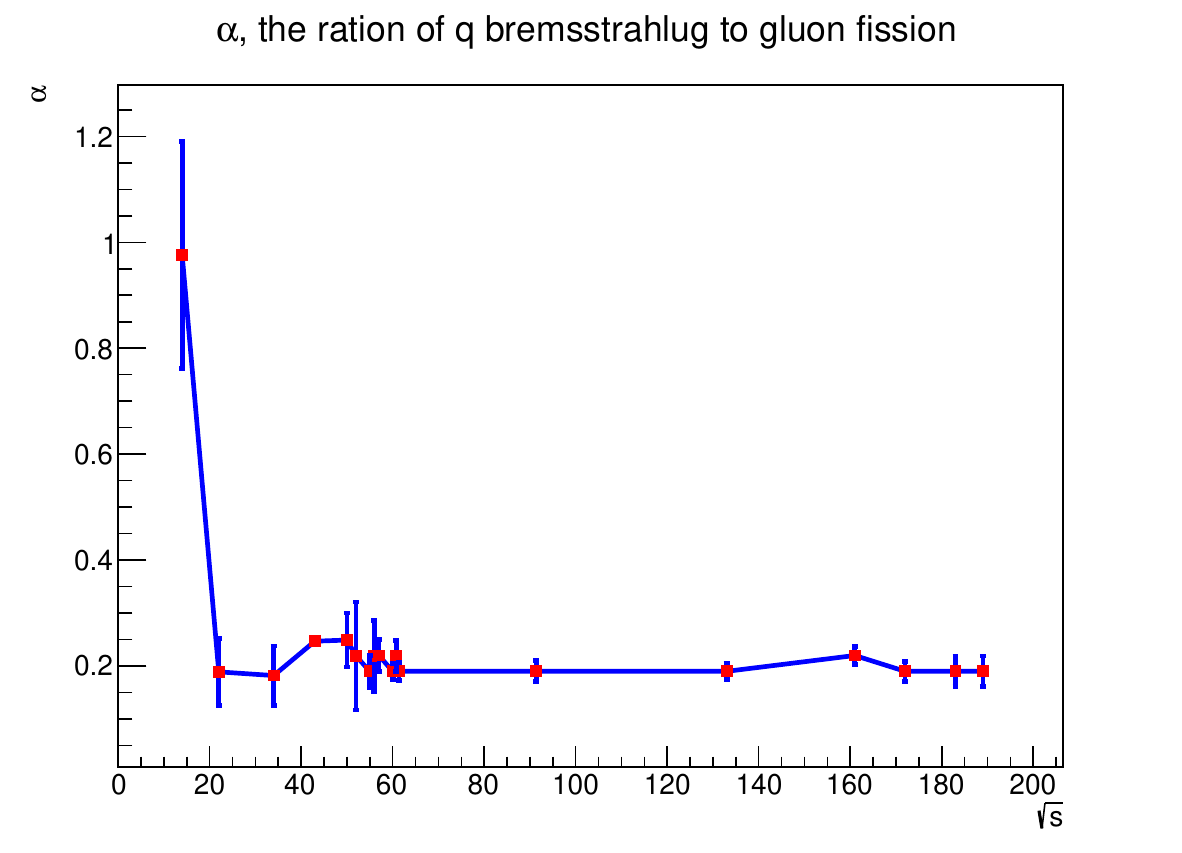}
    \includegraphics[angle=0, width=0.45\textwidth]{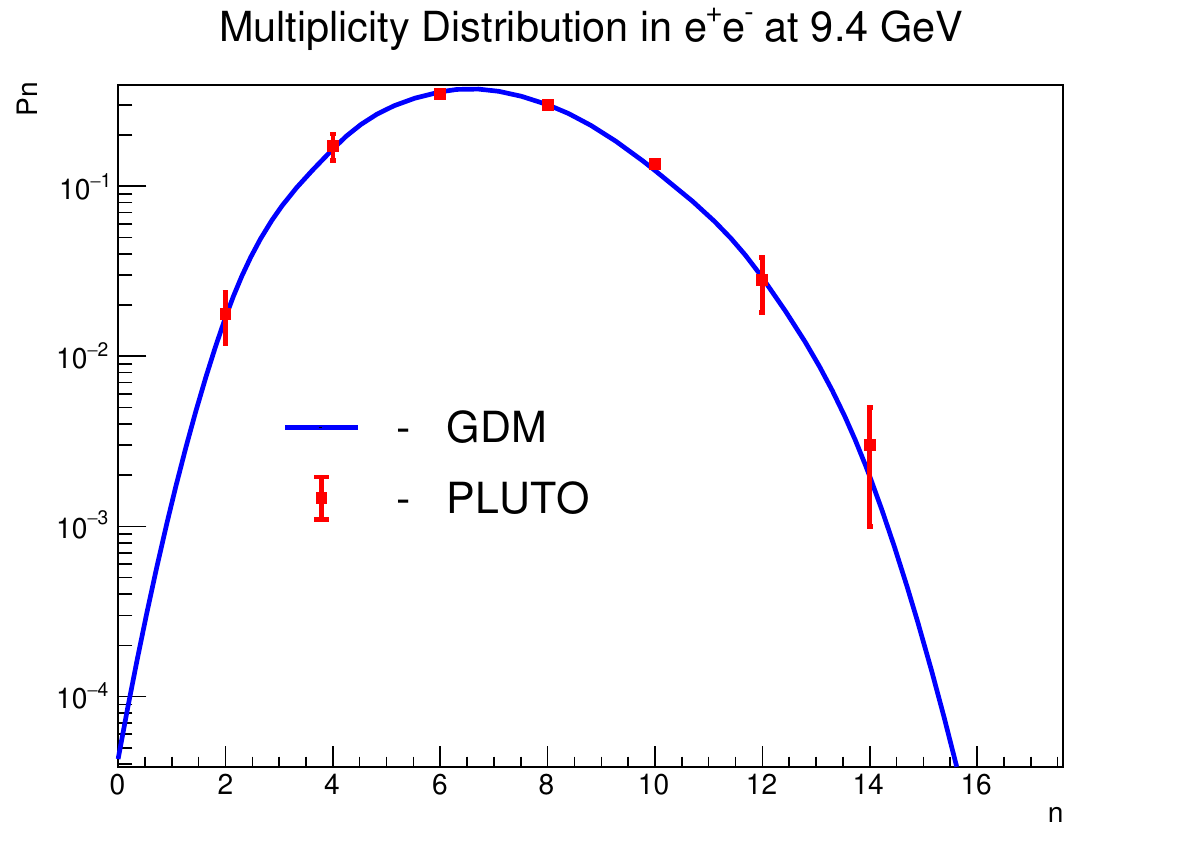}
	\caption{The GDM parameter, $\alpha$ (on the left), the description of MD in $e^+e^-$ annihilation by GDM and data at 9.4 GeV \cite{PLUTO} (on the right).} 
	\label{fig12}
\end{figure}

The parameter $\alpha $ is also remains almost constant excluding 14 GeV (Figure \ref{fig12}, on the left). This energy is the threshold between predominance of a q-bremsstrahlung at low energy over gluon fission and the notable participation of it in MP at higher energy.

We now apply the GDM for the description of MD at 9.4 GeV. Unfortunately, the data at 9.4 GeV were taken from \cite{PLUTO} where they were presented in a logarithmic scale as a KNO function. Therefore, obtaining their precise values was difficult. Excluding of gluon fission leads us to Poisson distribution for gluon due to q-bremsstrahlung \cite{Giov}. MD of charged hadrons in this case corresponds to

\begin{equation}
\label{eq24}
P_m(s) = \sum _{m=0}^{M_g} \frac{\bar m^m e^{-\bar m}}{m!} \binom{(2+\alpha m)N}{n}
\left (\frac{\overline n ^h}{N}\right ) ^n \left ( 1-\frac{\overline n^h}{N}\right )^{(2+\alpha m)N-n}.
\end{equation}
The second correlative moment for this MD is equal to 
\begin{equation}
\label{eq25}
f_2 = \left ({\alpha }^2 \bar m - \frac{2+\alpha \bar m}{N} \right ) (\bar {n}^h)^2.
\end{equation}

MD at this energy is shown in Figure \ref{fig12} (on the right). The maximum  number of gluons in the sum (\ref{eq24}), MG is equal 7. GDM predicts maximum number of hadrons 14, its parameters are equal: $\overline n^h_g$ = 0.97 $\pm $ 0.51 (that again confirms the fragmentation mechanism of hadronization). The average multiplicity of gluons is equal to 3.22 $\pm $ 0.99, $\bar n^h$ =1.93 $\pm $ 0.38, $\alpha $ = 0.5 $\pm $ 0.2, $N$ = 2.7 $\pm $ 1.4 and $f_2$ is negative.

Attempts using a complete accounting of the $qg$-cascade for our MD description by (\ref{eq19}) demonstrate the impossibility of determining the parameters at the cascade stage, while reasonably determining the hadronization parameters consistent with the previous values, in particular with the applying for cascade the formula (\ref{eq24}).

\section{Conclusion}
In this paper we have shown that GDM describes well MD in $e^+e^-$-annihilation by a two-stage scheme. This model relies on the convolution of two stages. The first stage, a quark-gluon cascade is described by a Markov branching process based on QCD. The second stage, hadronization uses phenomenological scheme based on data. This model uses six main parameters (including a normalization factor), which are determined directly from comparison with the data. 

Our findings allow drawing the reasonable picture of interaction. A q-bremsstrahlung is predominant at low energies less than 10 GeV when a $qg$-cascade is not developed enough. At that, a gluon fission is suppressed by roughly two orders of magnitude. With increasing of energy its contribution become notable and already makes up one fifth. The threshold of this crossing coincides with the zero value of $f_2$.

The gluon hadronization parameter, $\overline n^h_g$ (we define it indirectly) demonstrates a constant value and $\le $ 1 at $\sqrt s < \sim $ 100 GeV (before the region of a $Z^0$-boson creation). This behavior evidences fragmentation mechanism of hadronization that also confirms the local-hadron duality hypothesis (LoPHD) \cite{QCD1}. Then, we observe its weak growing, it becomes a little greater than 1 ($\sim $ 1.2). Such values are characteristic of the recombination mechanism of hadronization observed by us in proton collisions \cite{GDM3}. 

As energy increases, the number of active gluons significantly exceeds two quarks, which broadens and softens jets although the hadronization of a quark is more tough compared to that of a gluon. A gluon emitted by $q$-bremsstrahlung or produced due to fission carries significantly less energy  than a quark that has lost it. Therefore, the average multiplicity of hadrons formed from a quark exceeds that of a gluon by approximately five times (alpha 0.2). Also, the gluon fission is the source of high multiplicity. 

The average charged multiplicity is determined by the expression, $\bar n(s)$ = $(2+\alpha \bar m) \bar n^h_q$.  To obtain its prediction at 500 GeV or 1 TeV, the fitting parameters $\bar m$ and $\bar n^h_q$ should be described by reasonable functions. We choose possible dependencies for every variable, $\bar m$ and $\bar n ^h_q$. 

\begin{figure}[H] 
	\leavevmode
	\centering	
	\includegraphics[angle=0, width=0.45\textwidth]{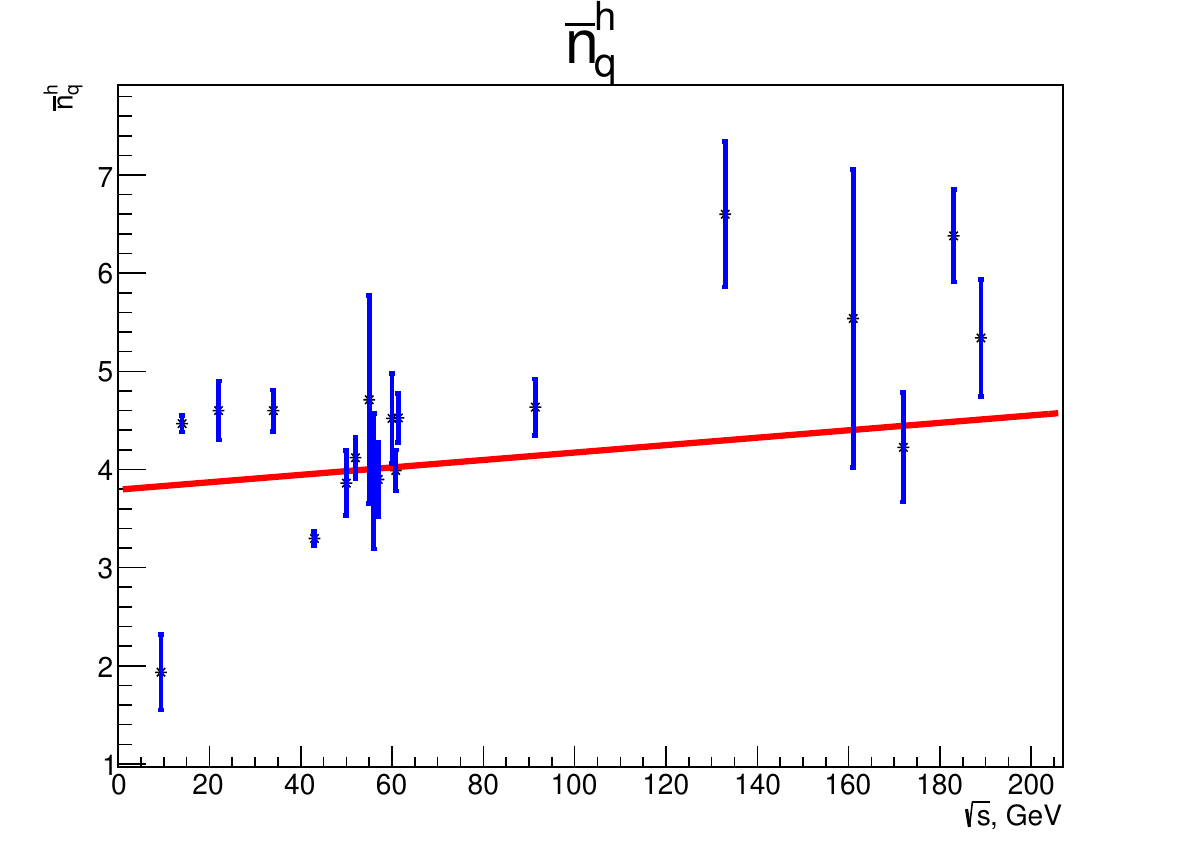}
	\includegraphics[angle=0, width=0.45\textwidth]{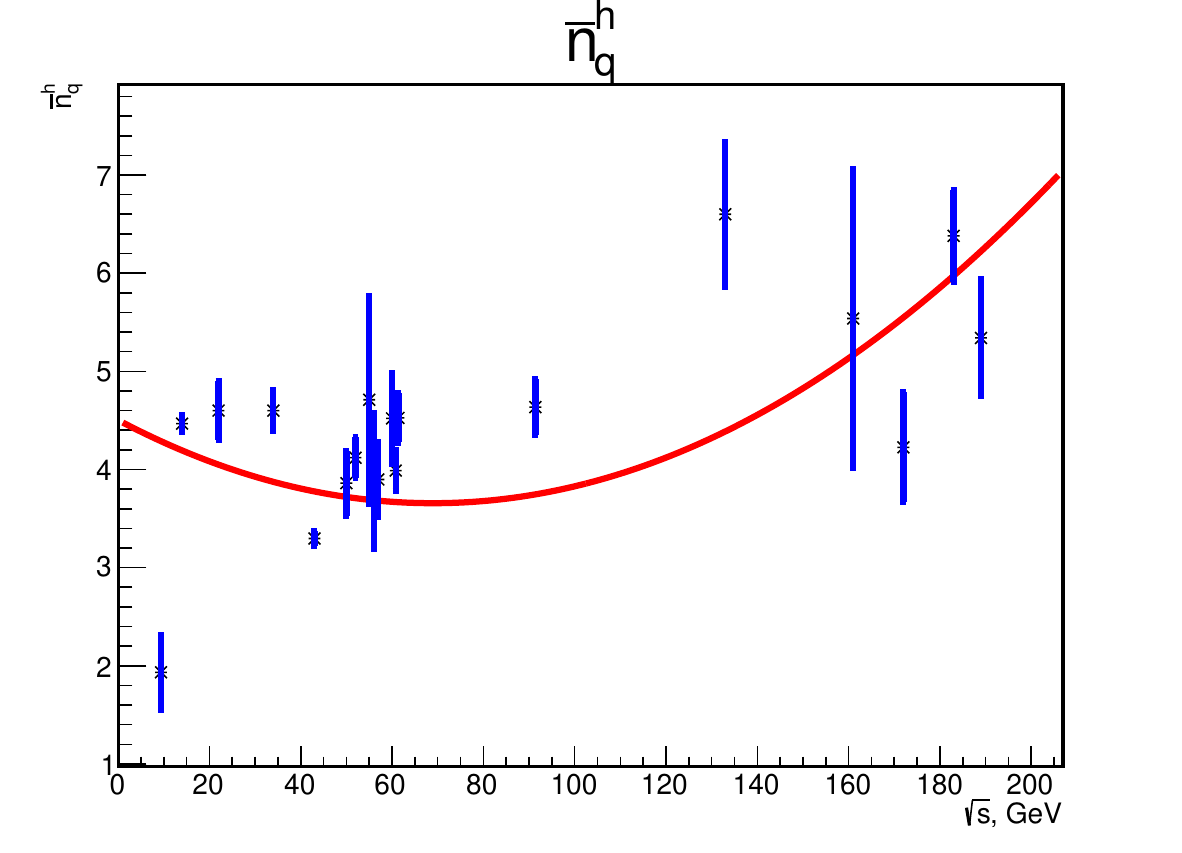}
	\caption{The description of linear growth with energy of a parameter $\bar n^h_q$  (on the left), the quadratic growth with energy of a parameter $\bar n^h_q$ (on the right).} 
	\label{fig14}
\end{figure}

The average number of hadrons formed from a single quark at hadronization, $\bar n^h_q$ is described by two functions and they are shown in Figure \ref{fig14}: the linear function $p_0+p_1 \sqrt s$, on the left, and the quadratic dependence of energy ($p_0 +p_1 \sqrt s + p_2 s$), on the right.

\begin{figure}[H] 
	\leavevmode
	\centering	
	\includegraphics[angle=0, width=0.45\textwidth]{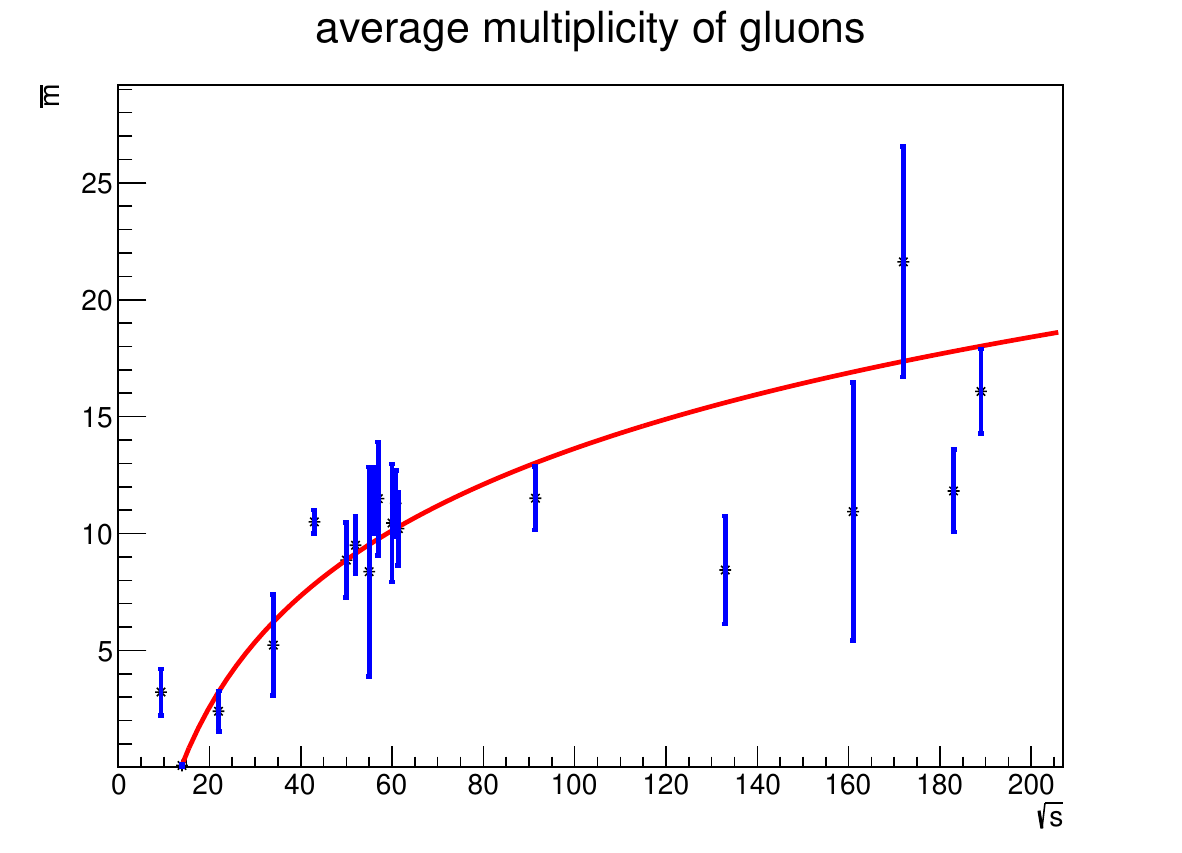}
	\includegraphics[angle=0, width=0.45\textwidth]{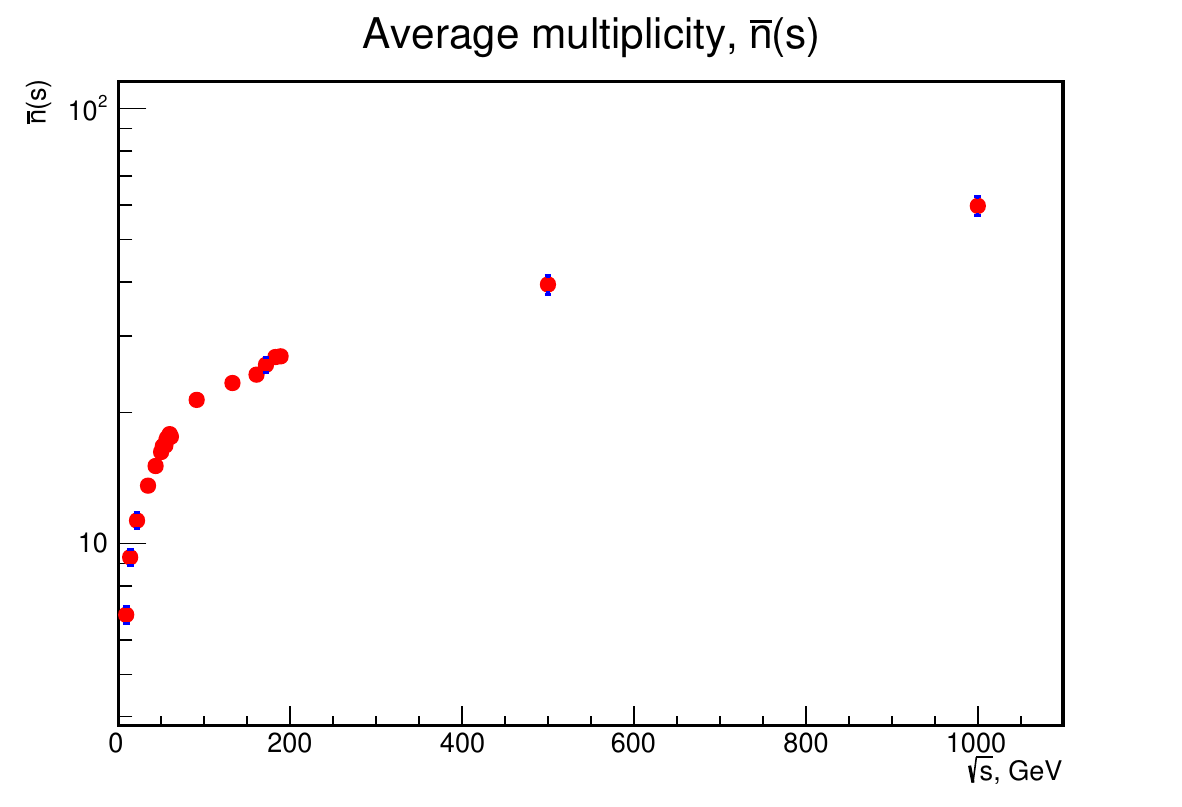}
	\caption{The description of average gluon multiplicity, $\overline m$  by logarithmic function, $p_0 + p_1 \cdot log s$  (on the left), the prediction of $\bar n(s)$ in the cases of: the logarithmic growth of $\bar m$ and linear for $\bar n^h_q$  (on the right). The last two top dots on the right belong for average multiplicity predictions at 500 GeV and 1 TeV } 
	\label{fig15}
\end{figure}

In Figure \ref{fig15} $\bar m$ is described by logarithmic function, $p_0+p_1 \log s$ (on the left). Using the explicit expression for the average number of gluons obtained by Giovannini (\ref{eq23}) \cite{Giov} is difficult because the energy dependence of the parameters $\tilde A$ and $A$ is not known.

We assume that $\alpha $ parameter remains approximately constant and equal to 0.2. 
By performing the possible combinations with $\bar m$ and $\bar n ^h_q$ (two permutations) one obtain the expecting values of $\bar n(s)$. In Figure \ref{fig15} one of such predictions is presented. The possible values of $\bar n(s)$ are varied for them from 32 to 39 at 500 GeV and from 37 to 60 at 1 TeV. In our opinion, both predictions look reasonable. 

The observed jumps in the behavior almost of all GDM parameters can be related with a  quark topology change (for example, the intermediate resonance creation) when the energy of colliding particles increases. On the contrary, we observe the absence of a expecting smooth transition for them. While the description in this model MD in proton-antiproton annihilation \cite{NPCS24} or in the heavy quarkonium decays \cite{Tomsk} we modified this topology

These and other findings obtained in GDM is causing increased interest to MP study.
The proposed model approach, along with Monte Carlo generators, allows one to make predictions of multiplicity and obtain, directly from comparison with the data, useful and important information.

The author expresses sincere gratitude to SVD-2 Collaboration for their significant contribution to MP study. We also hope that the study of multiparticle production will be continued in the SPD project, since it will have significantly greater capabilities for this.
\begin{center}
	
\end{center}

\label{last}
\end{document}